\documentclass[%
reprint,
 amsmath,amssymb,
pra,
]{revtex4-2}

\usepackage{graphicx}% Include figure files
\usepackage{dcolumn}% Align table columns on decimal point
\usepackage{bm}% bold math
\usepackage{physics}
\usepackage{amsmath}
\usepackage{amssymb}
\usepackage{slashed}
\usepackage{hyperref}
\usepackage{xcolor}
\usepackage[utf8]{inputenc}

\begin{document}

%\preprint{APS/123-QED}

\title{1D Photonic Band Gap Atlas, Formula Extension and Design Applications}

\author{Oscar D. H. Pardo}
 \email{osdhernandezpa@unal.edu.co}
\author{R. R. Rey-Gonz\'alez}
 \email{rrreyg@unal.edu.co}
\affiliation{
Universidad Nacional de Colombia,
Facultad de Ciencias, 
Departamento de F\'isica,
Grupo de \'Optica e Informaci\'on Cu\'antica\\
Grupo de Materiales Nanoestructurados y sus Aplicaciones,
Ciudad Universitaria 111321 Bogot\'a D.C. Colombia
}

\date{\today}

\begin{abstract}
The design and development of new photonic devices for technological applications requires a deep understanding of the effect of structural properties on the resulting band gap size and its position. Here, we perform a theoretical study of behavior of the photonic band gap sizes, positions and percentages under variations of the parameters characterizing binary (two materials), ternary (three materials) and linear dielectric grating multilayer structures. The resulting band gap atlas show that binary systems may suffice for most applications but ternary systems may add additional flexibility in design if needed. Linear gratings show a regular pattern for all gaps studied, this regularity was able to be reproduced with only few materials involved. The position of the gaps showed a very monotonous behavior for all calculations performed. Finally, additional extensions of formulas commonly used in the design of Bragg mirrors/reflectors using binary materials were proposed with their corresponding limitations discussed. These results can be seen as a technological horizon for photonic device development.

\begin{description}
\item[Keywords] 
Photonic Crystals, Photonic Band Gap Atlas, Multi-layer Systems. 
\end{description}
\end{abstract}

%\keywords{Photonic Crystals, Photonic Band Gap Atlas, Multi-layer Systems}

\maketitle

\section{\label{sec:intro}Introduction}

Photonic crystals (PCs) are systems which offer both plenty of intriguing natural phenomena \cite{labradorita, camaleones-pc-naturaleza, mariposa-naturaleza} and a ever growing list of real world applications \cite{review-inspirados, review-naturaleza-aplicaciones, review-polimeros-avances-recientes}. Even, while focusing on 1D PC, there is still plenty of approaches to obtain a wide variety of physical effects \cite{review-polimeros-avances-recientes, control-polimeros-espejos-filtros, review-cnc-polimeros, review-pc-deposicion-liquida, review-polimeros-mas-contraste}. The main applications of 1D PCs are as mirrors \cite{espejos-inkjet-formulas, espejos-binario-ejemplos} or filters \cite{filtro-binario-1, filtro-binario-2}, but there are also several growing uses in areas like solar cell research \cite{review-celdas-solares, celdas-perovskitas-1, celdas-perovskitas-2} or optical cavities like those used in quantum optics \cite{review-cavidades-opticas, pc-trampas-atomicas-cavidades, cavidad-optica-punto-q}. There are several applications where dynamic color change is utilized, like in sensors \cite{review-sensores-hidrogeno, review-sensores-gases-toxicos, review-biosensor-2012, sensor-glucosa-binario, sensor-glucosa-ternario}, smart skins, \cite{smart-skin-cefalopodos-inspirado} and angular filters, \cite{filtro-angular-1, filtro-angular-2, filtro-angular-3} with applications in telecommunications.

Mapping the PC band gaps when the parameters that define it are changed, is a great way to visualize the possibilities that a specific structure may offer, both in terms of application and physical properties. This often goes by the name of ``atlas of band gaps" \cite{joannopoulos, atlas-3d, atlas-2d-3d}. The mappings are usually only done for 2D or 3D structures. Unlike 2D and 3D PCs, 1D PCs tend to have several photonic band gaps, and these tend to exist for most of the possible parameter values that a PC structure may have. In this work, for the first time, we provide a variation of the ``atlas of band gaps" concept for the 1D PC case. These mappings offer important information in cases where the position and size of the band gap is desired to change dynamically \cite{cnc-dinamico-inspirado, dinamico-cavidades, dinamico-voltajes, dinamico-contacto-inspirado}.

On the other hand, in the design of mirrors or optical filters formulas for average frequency, gap position ($\overline{\omega}$) and ratio between the photonic gap size and the average frequency, percentage estimation, ($\Delta \omega / \overline{\omega}$) \cite{joannopoulos, espejos-inkjet-formulas, celdas-perovskitas-1, espejos-binario-ejemplos, cavidad-optica-punto-q, filtro-angular-3} are used. However, these formulas are only valid for the first spectral gap in the specific case of a quater wave stack structure. Since the quater wave stack maximizes the band gap (the gap percentage) \cite{demostracion-cuarto-onda} this offers the ideal structure for many static 1D application utilising the first spectral gap. Additionally, gap position estimation can also be provided by effective medium approximation (EMA) models (or flavours) \cite{ema-teorico, berthier, review-cavidades-opticas, review-polimeros-avances-recientes, espejos-inkjet-formulas, teorico-binario-fdtd}.

We focus on the possibilities offered by the binary 1D PC to control photonic band gaps and on the options introduced by adding more layers to the PC. Specifically, we focus on the ternary and linear dielectric grating 1D PCs. For all systems, the applicability for gap position estimation of the different EMA approximations is studied. For the binary case, an ``atlas of photonic band gaps'' is proposed, the emerging patterns are studied and the extensibility of gap position and percentage estimation formulas is also considered, even including higher order gaps. The limitations of this extensions are also discussed. For the ternary case, the role of each of the parameters is described qualitatively. Finally, the linear dielectric grating case showed interesting properties that where attempted to be reproduced with low numbers of discrete materials. This case also showed anomalous behavior of the gap position, which was resolved by proposing an alternative for the gap estimation formulas. These results have the potential to improve the design of dielectric structures by allowing the exploration of a vast number of possible structures with minimal computational cost. 

\section{\label{sec:marco-t}Theoretical Framework}

\subsection{Physical Model}

PCs are periodic arrays of dielectric function variations. Just like the appearance of band gaps due to periodicity in the potential in solid state crystals may prevent the occupation of energy levels by electrons, the periodicity in dielectric function can induce band gaps in electromagnetic (EM) radiation dispersion relations that prohibit light propagation with certain energies (and therefore frequencies) \cite{joannopoulos}. In particular, planar 1D PCs always have degenerate transverse electric (TE) and transverse magnetic (TM) modes, which means they offer the same gap structure under normal incidence. This is not necessarily the case in 2D or 3D PCs.

The reflection mechanism of PCs results in a mirror or filter effect which may offer a greater intensity, than the one obtained through other means \cite{berthier}, for the bandwidth the crystal is designed for.

Although, the band structure may offer important information beyond the gap size and position (like group velocities or density of states), in this case, we are interested on the band gap information of the band structures of the PC in question. For this reason we focus on the gap size ($\Delta \omega$), gap position ($\overline{\omega}$) and gap percentage ($\Delta \omega / \overline{\omega} $).
The gap position is related to the average of the frequencies of light that may be reflected from the crystal, whereas the gap size is related to the bandwidth of frequencies that may be reflected by the crystal.

\subsubsection{Binary system}

\begin{figure}[ht!]
    \centering
    \includegraphics[scale=0.45]{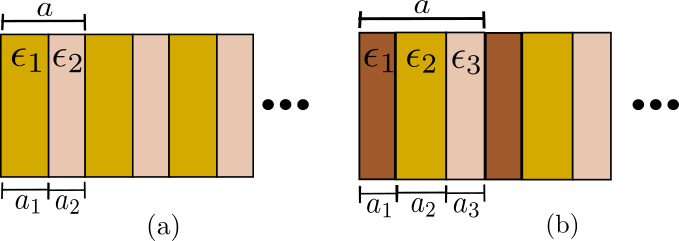}
    \caption{ Visual representation of a binary (a) and ternary (b) PCs, with the parameters that characterize each one}
    \label{fig:binario-ternario}
\end{figure}

The simplest 1D PC is a binary PC. It is composed of a repeating pattern of two blocks of different dielectric constants (figure \ref{fig:binario-ternario} (a)). The direction of repeating dielectric pattern is the one that defines the 1-dimensional (1D) PC.
We can parameterize all binary 1D PCs using 4 parameters: The two dielectric constants of the materials ($\epsilon_1$,$\epsilon_2$) and the thickness of each material ($a_1$, $a_2$), where $a=a_1 + a_2 $ is the lattice parameter of the crystal. As a matter of convention, we also choose $\epsilon_1 > \epsilon_2$.

Now, we actually are interested mainly in two of the parameters: The dielectric constant of the material with larger dielectric constant ($\epsilon_1$) and the size of the material with larger dielectric constant ($a_1$). This is because we may normalize the lattice parameter as $a=1$, and thus, $a_2 = 1 - a_1$ would be determined in this unit system. After doing the band structure calculations the results may be scaled back to any unit system of choice. Yet, that goes beyond the scope of this work, so the normalized units are enough for us. Also, we may set the dielectric constant of one material to 1 as well. In this case, we may also scale to other values of the dielectric constant by using the scaling relations \cite{joannopoulos}(pag. 21-22)
\begin{eqnarray}
    \epsilon'(x) = \frac{\epsilon (x)}{s^2} \quad ,  \label{ec:escala-epsilon} \\
    \omega' = s \omega \quad . \label{ec:escala-omega}
\end{eqnarray}
Where $\epsilon(x)$ is the dielectric function defined over the primitive cell, and $s$ is an arbitrary scaling factor.
Thus, we may obtain a equivalent system 
\begin{equation}
    \epsilon' (x) =
    \begin{cases}
        \frac{\epsilon_1}{\epsilon_2} \quad , \quad \text{if $0 < x < a_1$} \quad ,\\
        1 \quad , \quad \text{ if $a_1 < x < 1$} \quad ,\\
    \end{cases}
\end{equation}
where the resulting frequencies will be related by $\omega'= \sqrt{\epsilon_2}\omega$. We, therefore, restrict out attention to the parameters $\epsilon_1$ and $a_1$. 

\subsubsection{Ternary system}

Ternary 1D PCs are composed of 3 different materials with different dielectric constants and sizes (figure \ref{fig:binario-ternario}(b)). Therefore, it requires 6 parameters ($\epsilon_1$, $\epsilon_2$, $\epsilon_3$, $a_1$, $a_2$, $a_3$) to fully characterize the system. But again, we will focus on 4 of the parameters ($\epsilon_1$, $\epsilon_2$, $a_1$, $a_2$) by taking advantage of the scaling properties seen in the previous section.

Furthermore, the ordering of the materials does not have an effect in the band structure since any ordering can be obtained from any other using a constant spatial displacement (which corresponds to a different choice of motive for the primitive cell) or a mirror transformation (which corresponds to a opposite convention for light propagation directions). In both cases, the physics remain invariant.
Thus, we can additionally use a convention where $\epsilon_1>\epsilon_2>\epsilon_3=1$.

\subsubsection{Linear dielectric grating system}

Lastly, a material with a linear dielectric grating would have a dielectric function defined by:
\begin{equation}
    \epsilon (x) = \epsilon_{min} + (\epsilon_{max}-\epsilon_{min})x \quad .
\end{equation}
In this case, there are only two parameters that characterize the material, since there are not any discrete step variations that defines the material (as in n-layer materials), but a continuous variation of dielectric grating. If, we use the convention $\epsilon_{min}=1$, then
\begin{equation}
    \epsilon (x) = 1 + (\epsilon_{max}-1)x \quad ,
\end{equation}
and we only need to focus on a single parameter to vary.

\subsection{Computational Details}

Photonic band calculations were performed using the MPB \cite{mpb} (MIT Photonic Bands) software, a free open source option implementing the plane wave method for photonic band calculations. It is possible to use this method because the equation that governs the properties of the PC (the master equation \cite{joannopoulos}) can be written as a eigen-value problem
\begin{equation}
    \nabla \times \bigg ( \frac{1}{\epsilon(\mathbf{r})} \nabla \times \mathbf{H}(\mathbf{r}) \bigg ) = \bigg (  \frac{\omega}{c} \bigg )^2 \mathbf{H}(\mathbf{r}) \quad ,
\end{equation}
and thus, the numerical problem is greatly simplified. Some advantages of frequency domain methods are the speed of the calculations, yet the scope of these calculations is usually smaller than for time domain simulations.

For most calculations, we were interested in the behavior of the first 5 gaps (for the systems described above). To that end, the calculations were performed using 20 bands, a resolution of 512, and mesh size of 50. The two lattice points of interest were the two boundaries of the 1-dimensional irreducible Brillouin zone (IBZ). We interpolated over 500 points in between these boundaries for the calculations. Required increments were performed, in the cases where additional number of calculated bands, resolution, mesh size or reciprocal lattice points were needed.

\section{\label{sec:resul}Results}

\subsection{Effects on a binary system}

Since the focus on the binary system is on the effect on the photonic band gaps of varying the parameters $\epsilon_1$ (dielectric constant of the material with higher dielectric constant) and $a_1$ (the relative size of the material with higher dielectric constant) on the system, we calculated the band structure and extracted the gap information for variations in the entire range of $a_1 \in (0,1)$ and a range of $\epsilon_1 \in (1,9.5]$.

\begin{figure}[ht!]
    \centering
    \includegraphics[scale=0.43]{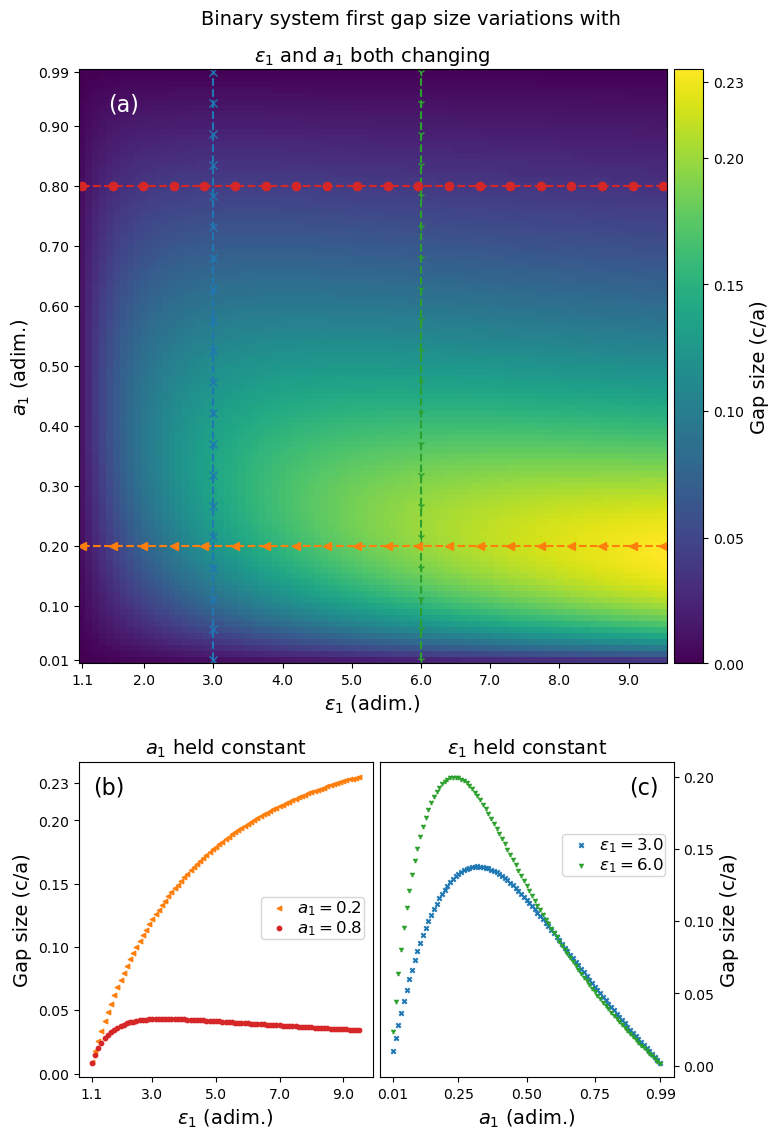}
    \caption{ Variations on a binary system of the size of the first gap (a) including both $\epsilon_1$ and $a_1$ variations (b) varying $\epsilon_1$ but holding $a_1$ in a fixed value (c) varying $a_1$ but holding $\epsilon_1$ in a fixed value.}
    \label{fig:2mat-gap1}
\end{figure}

\begin{figure}[ht!]
    \centering
    \includegraphics[scale=0.43]{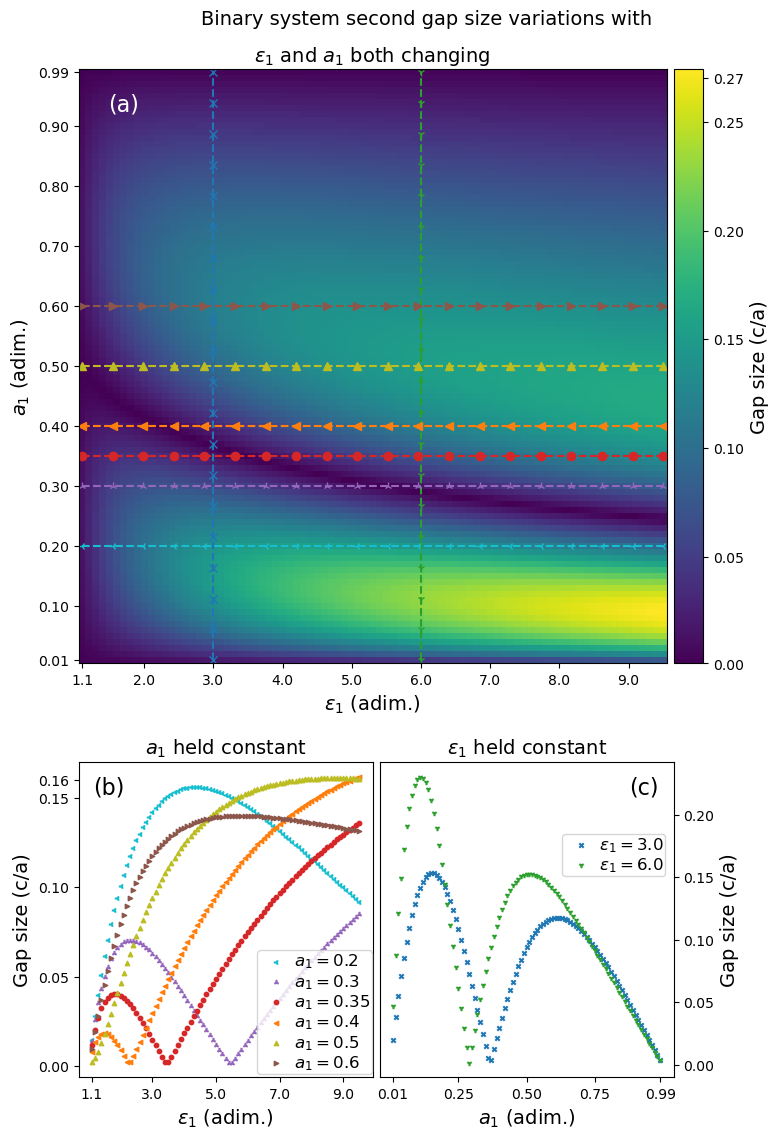}
    \caption{ Variations on a binary system of the size of the second gap (a) including both $\epsilon_1$ and $a_1$ variations (b) varying $\epsilon_1$ but holding $a_1$ in a fixed value (c) varying $a_1$ but holding $\epsilon_1$ in a fixed value. }
    \label{fig:2mat-gap2}
\end{figure}

\begin{figure}[ht!]
    \centering
    \includegraphics[scale=0.47]{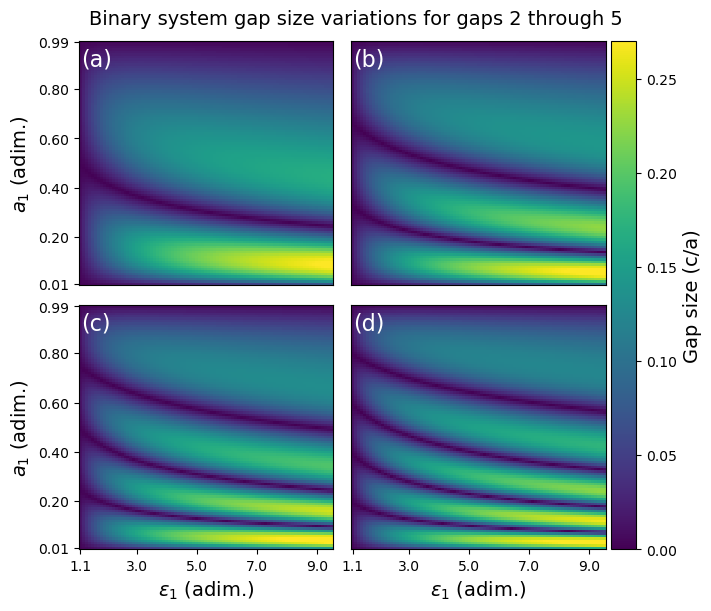}
    \caption{  Variations on a binary system of the size of the (a) second, (b) third, (c) fourth, (d) fifth gaps when both $\epsilon_1$ and $a_1$ are varied.}
    \label{fig:2mat-4gaps}
\end{figure}

The effects show distinct behaviors for each gap number, as can be seen in figures \ref{fig:2mat-gap1}, \ref{fig:2mat-gap2} and \ref{fig:2mat-4gaps}. We focus on both the global effect of varying both parameters and the effect of varying one parameter while holding the second fixed (projections). The variations show a number of ``hills" equal to the gap number and a number of ``valleys" equal to the gap number minus one. By ``hill" is meant the zones appearing in green to yellow in the color map figures (figures \ref{fig:2mat-gap1}(a), \ref{fig:2mat-gap2}(a), \ref{fig:2mat-4gaps}) and by ``valley" is meant the deep blue regions in between the ``hill" regions, which correspond to local extreme values. This description on terms of ``hills" and ``valleys" seemed the simplest way to capture the general features of the variations without creating confusion with the following discussion on the local maxima and minima in the projections. Thus, the term ``hill" and ``valley" will be restricted to discussion about two parameter variations whereas local minima and maxima will be used for the projections. As will be seen later, each of the ``hills" has a clearly defined ``ridge". The ``valleys"  and ``ridges" follow  a distinct pattern.

Now, focusing on the first gap, we see that the maximum values for the gap size occur for relatively low values of $a_1$ but for high a $\epsilon_1$ as possible (figure \ref{fig:2mat-gap1}(a)). Also, varying $\epsilon_1$ while keeping $a_1$ fixed (figure \ref{fig:2mat-gap1}(b)) shows a monotonically increasing gap size for increasing $\epsilon_1$ when $a_1$ is relatively low (figure \ref{fig:2mat-gap1}(b)(orange)). However, a local minima can appear for high enough values of $a_1$ (figure \ref{fig:2mat-gap1}(b)(red)). Although the overall scale in figure \ref{fig:2mat-gap1}(b)(red) is smaller, the local maxima can appear for lower values of $a_1$. On the other hand, varying $a_1$ while keeping $\epsilon_1$ fixed (figure \ref{fig:2mat-gap1}(c)) shows a similar behavior for all values of $\epsilon_1$. The effect of choosing a different value of $\epsilon_1$ simply moves the maxima seen in (figure \ref{fig:2mat-gap1}(c)) and scales the overall size of the gap. Although this effects may seem simple, they also form the basis for making sense of the more complex behavior of the other gaps.

The effects on the second gap are related to the effects on the first gap but show added complexity. The clearest difference is the fact that, whereas a single ``hill" appeared on the variations for the first gap (figure \ref{fig:2mat-gap1}(a)), two ``hills" appear for the second gap (figure \ref{fig:2mat-gap2}(a)). However, the way that the two ``hills" are separated have a profound effect in the projections. Specifically, varying $\epsilon_1$ while keeping $a_1$ fixed now has the possibility of having a local maxima for much smaller values of $a_1$, much closer to the highest gap sizes available to the system (figure \ref{fig:2mat-gap2}(b)(cyan)). Furthermore, an additional local minima can appear (figure \ref{fig:2mat-gap2}(b)(purple)) for slightly larger values of $a_1$. This local minima has a different position (in $\epsilon_1$) depending on the value of $a_1$; it appears at smaller values of $\epsilon_1$ as $a_1$ increases (figure \ref{fig:2mat-gap2}(b)(purple, red, orange)). For large enough values of $a_1$, the variations recover a similar behavior to the first gap, monotonically increasing or having one single local maxima with no local minima (figure \ref{fig:2mat-gap2}(b)(yellow, brown)) as in figure \ref{fig:2mat-gap1}(b). The reason for the behavior of the local minima is that the shape of the ``valley" that separates the two ``hills" is not fully horizontal, as variations of $\epsilon_1$ keeping $a_1$ fixed try to capture (see horizontal lines in figure \ref{fig:2mat-gap2}(a) and compare to ``valley" shape). The functional shape of these ``valleys" will be discussed later.

The behavior of the variations of $a_1$ while keeping $\epsilon_1$ constant are also very different from the first gap (figure \ref{fig:2mat-gap2}(c)). In this case, two local maxima appear on the projection, whereas a single one existed for the first gap. The effect of varying $\epsilon_1$ simply changes the positions of the maxima and minima and the overall scale but keep the general behavior constant.

For the third gap, an additional local ``moving" minima appears for the variations of $\epsilon_1$ keeping $a_1$ fixed (although the two minima are not seen simultaneously for any value of $a_1$), and an additional local maxima appears for the variations of $a_1$ keeping $\epsilon_1$ fixed, as can maybe be seen from figure \ref{fig:2mat-4gaps}(b) and the previous explanation. In general (figure \ref{fig:2mat-4gaps}), a number of ``hills" appears, equal to the gap number, with a number of ``valleys" equal to the gap number minus 1. This results in a number of local maxima for variations of $a_1$ keeping $\epsilon_1$ fixed equal to the number of ``hills" (gap number), and a number of local ``moving" minima for variations of $\epsilon_1$ keeping $a_1$ fixed equal to the number of ``valleys" (gap number minus 1). The ``moving" local minima do not appear simultaneously for the gap numbers shown, but for high enough gap number (at least seven) several ``moving" local minima can be seen simultaneously for a fixed value of $a_1$.

Another general aspect of the gap size variation is that the maximum value for the gap size occurs for increasingly lower values of $a_1$ as the gap number increases (figure \ref{fig:2mat-4gaps}), but for as high an $\epsilon_1$ as can be achieved.

On the other hand, the gap percentage ($\Delta \omega / \overline{\omega}$) variations have a qualitatively similar behavior to the gap size variations. The differences will be mentioned below, when quantitative considerations are included.

\begin{figure}[ht!]
    \centering
    \includegraphics[scale=0.57]{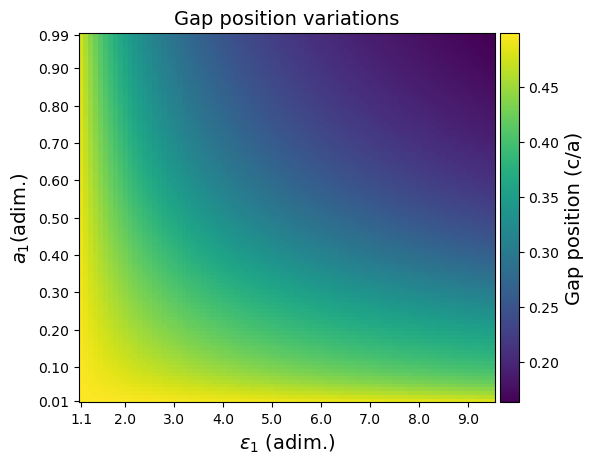}
    \caption{ Variations of the first gap position when both $\epsilon_1$ and $a_1$ are varied.}
    \label{fig:2mat-prom}
\end{figure}

Finally, aside from the gap size and percentage variations, the gap position variations are a lot more monotonous, as can be seen in figure \ref{fig:2mat-prom}.

\subsubsection{Effective Medium Approximation (EMA) - Drude and OPL flavours}

The simplicity of the position variation (figure \ref{fig:2mat-prom} ) suggests that an effective medium approximation (EMA) \cite{ema-teorico} could provide a good estimate for the gap position, as also suggested by the known equation (see for example \cite{teorico-binario-fdtd}):
\begin{equation}
    \label{ec:efectivo}
    \overline{\omega} = \frac{m}{2\sqrt{\epsilon_{eff}}} \quad .
\end{equation}
With $m$ an integer.

In the literature, there are two models or flavours from the EMA framework that appear repeatedly for binary PCs: a Drude \cite{berthier, review-cavidades-opticas} (Silberstein) model 
\begin{equation}
    \label{ec:drude}
    n_{drude}^2 = a_1 n_1^2 + a_2 n_2^2 \quad ;
\end{equation}
and a more popular optical path length (OPL) based, or ``parallel" (Birchak) flavour \cite{review-polimeros-avances-recientes, espejos-inkjet-formulas, teorico-binario-fdtd}:
\begin{equation}
    \label{ec:paralelo}
    n_{op} = a_1 n_1 + a_2 n_2 \quad .
\end{equation}
We may use this formulas to estimate the gap position by using the relation between the refractive index and dielectric constant ($n=\sqrt{\epsilon}$) and replacing (\ref{ec:drude}) or (\ref{ec:paralelo}) in (\ref{ec:efectivo}).

\begin{figure}[ht!]
    \centering
    \includegraphics[scale=0.48]{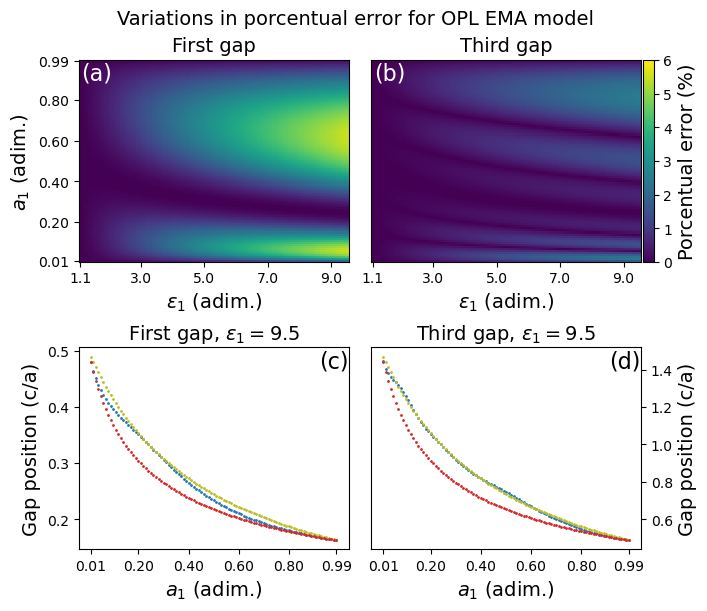}
    \caption{ Upper panels: Relative error evolution, into the OPL flavour, for position of the first (a) and third (b) gaps. Lower panels: Position of the first (c) and third (d) photonic gaps for Drude (red), Birchak (yellow) flavours and full simulation (blue).}
    \label{fig:error-ema}
\end{figure}

Even though both EMAs provide reasonably good estimates for the gap position, the Drude model works better close to the mono-layer limits ($a_1 \sim 0,1$), the OPL model works better overall, especially close to the regions that maximize the gap percentage. The relative error of the OPL model is shown in figure \ref{fig:error-ema}, for the first(a) and third (b) gaps. The error is small for low values of $\epsilon_1$ and it grows as $\epsilon_1$ does. For every gap number, there are areas where the error tends to zero, we will later connect these areas with the ``valleys" and ``ridges" of the corresponding gap percentage variations. The evolution of the position of the first and third photonic band gap is analyzed as function of $a_1$ parameter using both Drude and Birchak flavours. The results obtained from these flavours follow the behavior observed in the full simulation \footnote{The full simulation is referred as the solution of Maxwell's equations in the layered system considering the periodicity in the dielectric constant} as can we see in figure \ref{fig:error-ema}, where the c(d) panel corresponds to the evolution of the first (third) photonic gap; the red, yellow and blue lines are for the Drude flavour, the OPL flavour, and the full simulation, respectively.

\subsubsection{Beyond a quarter wave case}
Additionally, we can compare known analytical results for gap percentage and positions in binary systems \cite{joannopoulos} with our results. Former results are only valid for the quarter wave stack case, nevertheless, the resulting insights are fruitful in understanding the structure of the photonic band gap variations beyond the quarter wave stack case. We will first review the known formulas for the quarter wave stack case and how they compare to our results. Afterwards, we will show how the quarter wave stack condition may be generalized to understand the structure of gap variations beyond the first gap. 

Finally, we will study the extension of equations for quarter wave stack beyond both this case and the first photonic band gap.

First, the quarter wave stack condition may be expressed as (in our case $\epsilon_2 = 1$, $a_1+a_2=1$) 
\begin{equation}
    \label{ec:cond-cuarto-onda}
    a_1 = \frac{1}{1+\sqrt{\epsilon_1}} \quad ,
\end{equation}
making it easier to compare to our results. Whenever this relation is fulfilled, the formulas \cite{joannopoulos}
\begin{equation}
    \label{ec:posicion-joanopoulos}
    \overline{\omega} = \frac{n_1 + n_2}{4 n_1 n_2} = \frac{\sqrt{\epsilon_1} + 1}{4 \sqrt{\epsilon_1} } \quad ,
\end{equation}
and 
\begin{eqnarray}
    \label{ec:porcentaje-joanopoulos}
    \frac{\Delta \omega}{ \overline{\omega}} & = &\frac{4}{\pi} \arcsin{\bigg ( \frac{n_1 - n_2}{n_1 + n_2} \bigg )} \nonumber \\
    & = &\frac{4}{\pi} \arcsin{\bigg ( \frac{\sqrt{\epsilon_1} - 1}{\sqrt{\epsilon_1} + 1} \bigg )} \quad ,
\end{eqnarray}
should be valid for the first gap. Indeed, comparing the results expected from the formulas with the points in our calculation that more closely fit the quarter wave stack condition shows a very good match.

Furthermore, the condition (\ref{ec:cond-cuarto-onda}) does indeed maximize the gap percentage ($\Delta \omega/ \overline{\omega}$) for a given $\epsilon_1$, as expected from known results \cite{demostracion-cuarto-onda}. This means that the $\epsilon_1,a_1$ values that satisfy (\ref{ec:cond-cuarto-onda}) result in the highest gap percentage for a given $\epsilon_1$. However, the maximum gap size is achieved for slightly lower values of $a_1$, so the condition (\ref{ec:cond-cuarto-onda}) maximizes the resulting gap percentage but not the gap size of the first spectral gap.

Also, the $\epsilon_1,a_1$ values that satisfy the condition (\ref{ec:cond-cuarto-onda}) seem to be located in the ``valley" of the second band gap variations. In general, the $\epsilon_1,a_1$ values that satisfy the condition (\ref{ec:cond-cuarto-onda}) follow the shape of one of the ``valleys" for even gap numbers and follow the shape of one of the ``ridges" for odd gap numbers. 

Additionally, the $\epsilon_1,a_1$ values that satisfy the condition (\ref{ec:cond-cuarto-onda}) do maximize both the gap percentage and the gap size for odd gap numbers greater than one, except for the upper most ``ridge"; that one has the same sort of deviation that the gap percentage and size have for the first gap. The fact that the $\epsilon_1,a_1$ values for the quarter wave stack condition follow so well features in the gap percentage variations for several gap numbers lead us to believe that a similar functional form $a_1(\epsilon_1)$ could potentially follow the shape of all the main features of the band gap variations (the other ``valleys" and ``ridges"). 

We now focus on extending the quarter wave stack condition (\ref{ec:cond-cuarto-onda}). We first obtain the local minima and maxima for each value of $\epsilon_1$. By fitting this numerical curves to curves of the form 
\begin{equation}
    \label{ec:curvas-mu}
    a_1 = \frac{\mu}{\mu + \sqrt{\epsilon_1}} \quad ,
\end{equation}
where $\mu$ is a fitting parameter to be determined.

\begin{figure}[ht!]
    \centering
    \includegraphics[scale=0.58]{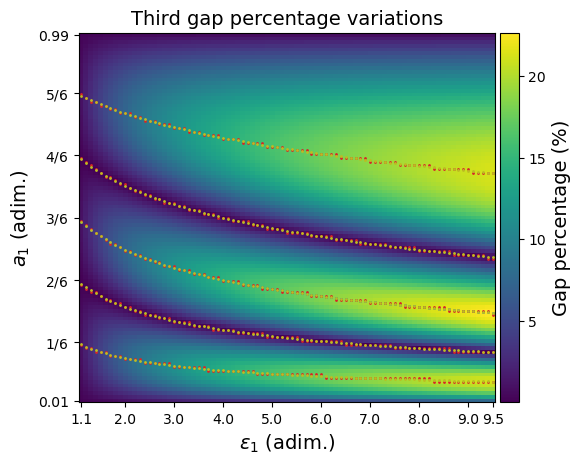}
    \caption{ Gap percentage variations for the third gap.  Numerically determined ``ridges" and ``valleys" (red curves). Curves determined using the tables (\ref{tab:mu-valles},\ref{tab:mu-crestas}) (yellow curves). The ticks for the y-axis have been chosen to match the limits of the aforementioned curves (see Eqs. 
    (\ref{ec:a*-crestas}) and (\ref{ec:a*-valles})).}
    \label{fig:anatomia}
\end{figure}

We found two separate patterns for the $\mu$ values for the ``valleys" and ``ridges". Figure (\ref{fig:anatomia}) shows the gap percentage variations of the third gap, over them, the $\epsilon_1$ and $a_1$ values that satisfy the numerically found extrema values for each $\epsilon_1$ (the ``ridges" and ``valleys"), and the corresponding curve of the form (\ref{ec:curvas-mu}) that better fits the corresponding ``ridge" or ``valley". 

\begin{table}[ht!]
    \centering
    \begin{tabular}{c|c}
        Gap number & $\mu$ values for ``valleys" \\
        \hline 
        2 & $\frac{1}{1}$ \\
        3 & $\frac{1}{2}$ \quad $\frac{2}{1}$ \\
        4 &  $\frac{1}{3}$ \quad  $\frac{2}{2}$ \quad  $\frac{3}{1}$ \\
        5 &  $\frac{1}{4}$ \quad  $\frac{2}{3}$ \quad  $\frac{3}{2}$ \quad  $\frac{4}{1}$  \\
        6 &  $\frac{1}{5}$ \quad  $\frac{2}{4}$ \quad  $\frac{3}{3}$ \quad  $\frac{4}{2}$ \quad  $\frac{5}{1}$ \\
        7 &  $\frac{1}{6}$ \quad  $\frac{2}{5}$ \quad  $\frac{3}{4}$ \quad  $\frac{4}{3}$ \quad  $\frac{5}{2}$ \quad  $\frac{6}{1}$ \\
        8 &  $\frac{1}{7}$ \quad  $\frac{2}{6}$ \quad  $\frac{3}{5}$ \quad  $\frac{4}{4}$ \quad  $\frac{5}{3}$ \quad  $\frac{6}{2}$ \quad  $\frac{7}{1}$ \quad \\
    \end{tabular}
    \caption{$\mu$ values for ``valleys" in gaps 2 trough 8. Replacing one of this values in the formula (\ref{ec:curvas-mu}) gives an equation for describing the corresponding "valley´´.} 
    \label{tab:mu-valles}
\end{table}

For the ``valleys" the $\mu$ values follow the pattern shown in table \ref{tab:mu-valles}.
This information can be summarized with the formula
\begin{equation}
    \label{ec:mu-valles}
    \mu_{i,m}^v = \frac{i}{m-i} \quad , \quad i = 1,...,m-1 \quad ,
\end{equation}
where $v$ corresponds to the fact that the formula applies to the description of ``valleys", $m$ is the number of the gap and $i$ is an integer. For this formula, the lowest $\mu$ value would correspond to the first ``valley" (from bottom-up) and the largest $\mu$ value would correspond to the last one (the uppermost ``valley").

\begin{table}[ht!]
    \centering
    \begin{tabular}{c|c}
        Gap number & $\mu$ values for ``ridges" \\
        \hline 
        1 & $\frac{1}{1}$ \\
        2 & $\frac{1}{3}$ \quad $\frac{3}{1}$ \\
        3 &  $\frac{1}{5}$ \quad  $\frac{3}{3}$ \quad  $\frac{5}{1}$ \\
        4 &  $\frac{1}{7}$ \quad  $\frac{3}{5}$ \quad  $\frac{5}{3}$ \quad  $\frac{7}{1}$  \\
        5 &  $\frac{1}{9}$ \quad  $\frac{3}{7}$ \quad  $\frac{5}{5}$ \quad  $\frac{7}{3}$ \quad  $\frac{9}{1}$ \\
        6 &  $\frac{1}{11}$ \quad  $\frac{3}{9}$ \quad  $\frac{5}{7}$ \quad  $\frac{7}{5}$ \quad  $\frac{9}{3}$ \quad  $\frac{11}{1}$ \\
        7 &  $\frac{1}{13}$ \quad  $\frac{3}{11}$ \quad  $\frac{5}{9}$ \quad  $\frac{7}{7}$ \quad  $\frac{9}{5}$ \quad  $\frac{11}{3}$ \quad  $\frac{13}{1}$ \\
        8 &  $\frac{1}{15}$ \quad  $\frac{3}{13}$ \quad  $\frac{5}{11}$ \quad  $\frac{7}{9}$ \quad  $\frac{9}{7}$ \quad  $\frac{11}{5}$ \quad  $\frac{13}{3}$ \quad  $\frac{15}{1}$ \\
    \end{tabular}
    \caption{$\mu$ values for ``ridges" in gaps 1 trough 8. Replacing one of this values in the formula (\ref{ec:curvas-mu}) gives an equation for describing the corresponding "ridge´´.}
    \label{tab:mu-crestas}
\end{table}

On the other hand, the $\mu$ values for the ``ridges" follow the pattern shown in table \ref{tab:mu-crestas}. This information can be summarized with the formula
\begin{equation}
    \label{ec:mu-crestas}
    \mu_{i,m}^r = \frac{2i+1}{2(m-i)-1} \quad , \quad i = 0,...,m-1 \quad ,
\end{equation}
where $c$ corresponds to the fact that the formula applies for the description of ``ridges" and $m$ is again the gap number. For this formula, the lowest $\mu$ value would correspond to the first ``ridge" (from bottom-up) and the largest $\mu$ value would correspond to the last one (the uppermost ``ridge").

Additionally, the entire reduced parameter space ($\epsilon_1$, $a_1$) can be filled with different values of $0 \leq \mu \leq \infty$, with $\mu = 0$ corresponding 
to $a_1^{0} = 0$ and $\mu \rightarrow \infty$, corresponding to $a_1^{\infty} = 1$. Thus, the curves defined by Eq. (\ref{ec:curvas-mu}) offer a way to understand the entire parameter space that naturally follows the structure in the gap size/percentage variations. Also, we can define the limit of any  $\mu$-curve as $\epsilon_1 \rightarrow 1$  like
\begin{equation}
    \label{ec:def-a*}
    a_1^{*} = \lim_{\epsilon_1 \to 1} a_1^{\mu} (\epsilon_1) = \lim_{\epsilon_1 \to 1} \frac{\mu}{\mu + \sqrt{\epsilon_1}} = \frac{\mu}{\mu + 1} \quad .
\end{equation}

With this limit we can obtain, for the specific cases of ``valleys" ($v$) and ``ridges"($r$):
\begin{eqnarray}
    a_{i,m}^{v*} = \frac{i}{m} \quad , \quad i = 1,...,m-1 \label{ec:a*-valles} \quad , \\
    a_{i,m}^{r*} = \frac{2i+1}{2m}  \quad , \quad i = 0,...,m-1   \label{ec:a*-crestas} \quad .
\end{eqnarray}

This implies an equal spacing for these limits on the y-axis, for both the spacing between two ``valleys" and the spacing between two ``ridges" (check figure \ref{fig:anatomia}),
\begin{equation}
    \label{ec:delta-escalera}
    \delta_m = a_{i+1,m}^{(v,c)*} - a_{i,m}^{(v,c)*} = \frac{1}{m} 
\end{equation}

Since the $\mu$ and $a_{\mu}^*$ values have a one-to-one correspondence ($\mu = a_{\mu}^*/(1-a_{\mu}^*)$), the $a_{\mu}^*$ not only serves as a way to characterize the different $\mu$ values, but also has a way to obtain the different $\mu_{i,m}^{v,c}$ allowed in a given gap number $m$. Notice that $\delta_m$ can be thought as a ``ladder operator" for the $\mu$ curves that correspond to crests/valleys. $\delta/2$ would go through both ``ridges" and ``valleys".

Having described the $\epsilon_1$ and $a_1$ values that make up the ``valleys" and ``ridges" of the band gap variations, we now turn to the extensibility of the formulas (\ref{ec:posicion-joanopoulos}, \ref{ec:porcentaje-joanopoulos}). First, we note that there is a big similarity between the equation for the $\mu$ curves (\ref{ec:curvas-mu})  and the quarter wave stack condition (\ref{ec:cond-cuarto-onda}) for $n_2 \neq 1$, where $\mu $ seems to replace $n_2$, the overall equation has the same functional form. However, $\mu$ and $n_2$ have very different ranges, so the similarity could very well just be a coincidence. Nevertheless, this does provide a possible avenue to extend the formulas (\ref{ec:posicion-joanopoulos},\ref{ec:porcentaje-joanopoulos}) beyond the quarter wave stack case.

For the position formula (equation \ref{ec:posicion-joanopoulos}) simply replacing $n_2$ with $\mu$
\begin{equation}
    \overline{\omega}_{\mu} = \frac{\mu + \sqrt{\epsilon_1}}{4\mu \sqrt{\epsilon_1}}
\end{equation}
does not work, yet noting that the condition 
\begin{equation}
    \label{ec:cond-posicion}
    \lim_{\epsilon_1,\mu \to 1} \overline{\omega}_{\mu} = \frac{1}{2}
\end{equation}
should be fulfilled for any $\mu$ value, the simplest extension found
\begin{equation}
    \label{ec:ext-posicion}
    \overline{\omega}_{\mu} = \frac{\mu + \sqrt{\epsilon_1}}{4(\mu+1) \sqrt{\epsilon_1}}
\end{equation}
gives a better estimate for the gap position. In fact, by replacing 
\begin{equation}
    \label{ec:mu-despejado}
    \mu = \frac{\sqrt{\epsilon_1}a_1}{1-a_1}
\end{equation}
we recover the estimate for the first gap given by the OPL EMA. 

On the other hand, the formula (\ref{ec:porcentaje-joanopoulos}) was not able to be generalized with any substitution of $n_2$. In fact, the functional form of (\ref{ec:porcentaje-joanopoulos}) already has the behavior seen for all $\mu$ values. Therefore, an extension of the form 
\begin{equation}
    \label{ec:ext-proporcionalidad}
    \frac{\Delta \omega}{\overline{\omega}} = \frac{4}{\pi} \nu (\mu)  \arcsin{\bigg ( \frac{\sqrt{\epsilon_1} - 1}{\sqrt{\epsilon_1} + 1} \bigg )} \quad ,
\end{equation}
where $\nu(\mu)$ is a proportionality constant with a specific value for each $\mu$, seemed the most appropriate. That is, the behavior of the gap percentage for any $\mu$ is proportional to (\ref{ec:porcentaje-joanopoulos}). Note that this proportionality is fulfilled along curves of constant $\mu$, that is, following the natural shape of the ``valleys" and ``ridges" in the band gap variations. This provides further incentive to think about the $\mu$- curves as the natural way to understand the band gap variations. Indeed, fitting the percentage values for each $\mu$ using the proportionality constant $\nu(\mu)$ as a fitting parameter for the first gap yields good results for the gap percentage. Later, we will see however that higher gaps deviate slightly from this behavior near the ``ridges".

For higher gap numbers, the $\nu$ function presents more complexity and the resolution quickly surpasses the capabilities of the calculations resolution. Nevertheless, $1/m$ does seem to provide an upper bound for $\nu$ for the m'th gap
\begin{equation}
    \label{ec:ext-cota}
    \frac{\Delta \omega}{\overline{\omega}_{\mu}} \leq \frac{4}{m\pi} \arcsin{ \bigg ( \frac{\sqrt{\epsilon_1} - 1}{\sqrt{\epsilon_1} + 1} \bigg ) } \quad .
\end{equation}

\begin{figure}[ht!]
    \centering
    \includegraphics[scale=0.47]{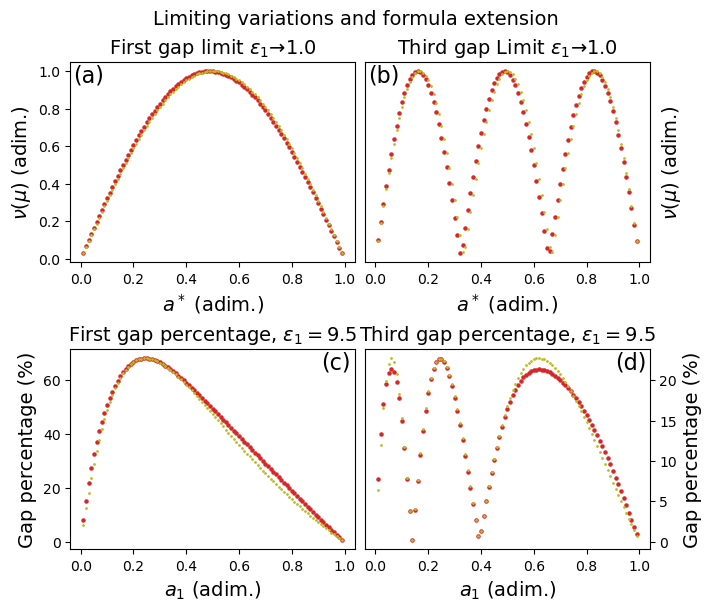}
    \caption{ Comparison between gap percentage variations in the small $\epsilon_1$ limit for the first (a) and third (b) gaps. Extension of the limit model to the rest of the reduced parameter space ($\epsilon_1$,$a_1$), we show the variations for $\epsilon_1=9.5$, since the relative error is greater there, for the first (c) and third (d) gaps. }
    \label{fig:extension}
\end{figure}

We can provide an estimate for the $\nu(\mu)$ function using the limiting behavior of the $a_1$ variations (keeping $\epsilon_1$ constant) as $\epsilon_1 \to 1$. This is because as may be seen in figures \ref{fig:2mat-gap1}(c),\ref{fig:2mat-gap2}(d) the complex structure of the gap size (or for that matter gap percentage) variations becomes simpler as $\epsilon \to 1$. Indeed, for $\epsilon_1 =1.1$ (the smallest value we considered) the variations are particularly simple (see figure \ref{fig:extension} (a),(b)). If we consider $\epsilon_1 =1.1$ as an approximation to the limiting case $\epsilon_1 \to 1$, which we take to resemble a limiting variation such that $\Delta \omega / \overline{\omega }= \Delta \omega / \overline{\omega } (a^*)$; then, we may obtain a function $\nu(\mu)$ by replacing $a^*(\mu)$ in the limiting variations. We may also replace $\mu(\epsilon_1,a_1)$ given by Eq. (\ref{ec:mu-despejado}) to obtain a final functional form in terms of $\epsilon_1$ and $a_1$. We explain this process in greater detail below.

We begin by choosing two candidates were considered to represent the limiting variations for the first gap, a quadratic and a sine like functions
\begin{eqnarray}
    \nu(a^*) = 4a^*(1-a^*) \quad , \label{ec:candidato-cuadratico} \\
    \label{ec:candidato-seno}
    \nu(a^*) = \sin{( \pi a^* )} \quad , \quad  a^* \in [0 , 1] \quad .
\end{eqnarray}

None require a fitting parameter but the sine function fits better, provides a better limiting behavior and has a straightforward generalization to greater gap numbers. Indeed, the general formula for the limiting variations for any gap is  simply given by (figure \ref{fig:extension} (a),(b))
\begin{equation}
    \label{ec:ext-a*}
    \nu(a^*) = \frac{|\sin{( m \pi a^* )}|}{m} \quad .
\end{equation}

Replacing in the gap percentage formula, equation \ref{ec:ext-proporcionalidad}, gives
\begin{equation}
    \label{ec:ext-mu}
    \frac{\Delta \omega}{\overline{\omega}}  = \frac{4 \big| \sin{ \big( \frac{m\pi \mu}{\mu+1} \big)} \big|}{m\pi} \arcsin{\bigg ( \frac{\sqrt{\epsilon_1} - 1}{\sqrt{\epsilon_1} + 1} \bigg )} \quad .
\end{equation}

This form may already prove useful, however, replacing $\mu$ can give the estimate for the gap percentage in terms of $\epsilon_1$ and $a_1$
\begin{equation}
    \label{ec:ext-formula}
    \frac{\Delta \omega}{\overline{\omega}} =  \frac{4 \big| \sin{ \big( \frac{m\pi a_1\sqrt{\epsilon_1}}{a_1(\sqrt{\epsilon_1}-1)+1} \big)} \big|}{m\pi} \arcsin{\bigg ( \frac{\sqrt{\epsilon_1} - 1}{\sqrt{\epsilon_1} + 1} \bigg )} \quad .
\end{equation}

This formula fits very well to the data, specially for small $\epsilon_1$ and the first gap (see figure \ref{fig:extension} (c)). For higher gaps, the formula still reproduces all the qualitative features of the variations but deviates significantly for some ``ridges". The further the $\mu$ value of the ``ridge" deviates from $\mu = 1$, the greater the deviation (see figure \ref{fig:extension} (c)(d)), so (\ref{ec:ext-formula}) still provides a very close estimate for ``ridges" with $\mu$ values close to $1$.

To understand the nature of the deviations from (\ref{ec:ext-formula}) further calculations were made, specifically along $\mu$ values corresponding to ``ridges". These calculations showed the following results. 

First, the OPL EMA model is exact along ``ridges". By exact, we mean that the results are as close as for the known case of (\ref{ec:posicion-joanopoulos}). This explains the ``valleys" in the relative error variations (figure \ref{fig:error-ema}(b)): OPL EMA is exact for both ``ridges" and ``valleys" (the valley case is exact since it corresponds to vanishing gaps). Notice that the third gap has three ``ridges" and two ``valleys" (see figure \ref{fig:anatomia} or (\ref{ec:mu-crestas}),(\ref{ec:mu-valles})) and the variations in relative error for the OPL EMA model has five ``valleys". Each of the ``ridges"/``valleys" for the corresponding gap (\ref{ec:mu-crestas}),(\ref{ec:mu-valles}) corresponds to a ``valley" in the relative error variations (figure \ref{fig:error-ema}).

Second, the way that each ``ridge" deviates from the formula (\ref{ec:ext-formula}) is a function of $\mu$ with the property $\nu'(\mu)=\nu'(1/\mu)$ irrespective of the gap number. That is, if two different gaps, both having a ``ridge" with the same $\mu$ value, then both ``ridges" have the same deviation from the formula (\ref{ec:ext-formula}). This observation only holds for ``ridges", away from these, the deviations follow more complex patterns.

A better approximation, exclusively near ``ridges" can be given for the gap percentage by
\begin{equation}
    \label{ec:ext-alt}
    \frac{\Delta \omega}{\overline{\omega}} \approx \frac{4 \big| \sin{ \big( \frac{m\pi \mu}{\mu+1} \big)} \big| e^{-0.0268\abs{\ln(\mu)}}}{m\pi} \arcsin{ \bigg( \frac{\sqrt{\epsilon_1} - 1}{\sqrt{\epsilon_1} + 1} \bigg) } \quad .
\end{equation}

The 0.0268 factor is an ad-hoc fit to the found deviations of the gap percentage formula, the exponential factor with the absolute value of the natural logarithm exponent is the simplest functional form that fulfills all the seen properties of the deviations seen for the different ``ridges". The estimate is better, specially around ``ridges" but worse for $a_1 \sim 0$ and $a_1 \sim 1$. There can be many alternative formulas which can provide a similar effect.

We quote the maximum relative error along ``ridges" using Eqs. (\ref{ec:ext-formula}) and (\ref{ec:ext-alt}) in table \ref{tab:error}. We also take the chance to quote the maximum relative error found for the OPL EMA flavour.

\begin{table}[h!]
    \centering
    \begin{tabular}{c|c|c|c}
          & \multicolumn{3}{c}{Maximum relative error estimating: } \\
          & \multicolumn{2}{c|}{Gap percentage along ``ridges"} & Gap position \\
        Gap number & without factor & with factor & (OPL) \\
        \hline 
        1 & $0.00 \% $ & $0.00 \% $ & $5.52\%$ \\
        2 & $3.79 \%$ & $0.78\%$ & $3.51\%$ \\
        3 & $6.52 \%$ & $2.03\%$ & $2.50\%$ \\
        4 & $8.11 \%$ & $2.61\%$ & $1.94\%$ \\
        5 & $9.15 \%$ & $2.91 \%$ & $1.58 \%$ \\
        6 & $9.85 \%$ & $3.01 \%$ & $1.33 \%$ \\
        7 & $10.33 \%$ & $3.00 \%$ & $1.14 \%$ \\
        8 & $10.86 \%$ & $3.11 \%$ & $1.01 \%$ \\
        9 & $11.10 \%$ & $2.98 \%$ & $0.90 \%$ \\
    \end{tabular}
    \caption{Maximum relative error achieved when estimating gap percentage (along ``ridges") and position using the formulas (\ref{ec:ext-formula}),(\ref{ec:ext-alt}),(\ref{ec:paralelo}) (respectively for each column) for each gap number. This error is achieved for the highest $\epsilon_1$ considered ($\epsilon_1 = 9.5$). Even higher $\epsilon_1$ is expected to yield higher error.}
    \label{tab:error}
\end{table}

Finally, the calculations along $\mu$ curves and the subsequent fitting showed that the increment along the ``ridges" deviated slightly from a $\arcsin$ behavior as was expected. The effect is non existent for $\mu=1$ and negligible for $\mu \sim 1$ but noticeable otherwise. This deviation from $\arcsin$ behavior is the main source of error quoted in table (\ref{tab:error}). It is worth noting that it is possible to combine (\ref{ec:efectivo}), (\ref{ec:paralelo}) and (\ref{ec:ext-formula}) to provide an estimate for the gap size. All the considerations we have outlined for the estimation of gap percentage also apply for the case of size estimation, including the use of formula (\ref{ec:ext-alt}).

Although the formulas described here ((\ref{ec:paralelo}),(\ref{ec:ext-formula}),(\ref{ec:ext-alt})) are only approximate, they still may facilitate the design of dielectric structures since the formulas allow the exploration of a wide arrange of possible configurations to be considered at a low computational cost. Also, the understanding of the gap size variations through the $\mu$ curve approach has the possibility to create the ground work for other systems beyond the binary dielectric one. 

\subsection{Effects on a ternary system}

For ternary systems, we focus on a parameter space with 4 parameters and, to gain insight in the role of each we break down the computations into three groups, each involving only two parameters. These groups are: ($\epsilon_1$, $a_1$) , ($\epsilon_1$, $\epsilon_2$) , ($a_1$ , $a_2$).

\begin{figure}[ht!]
    \centering
    \includegraphics[scale=0.43]{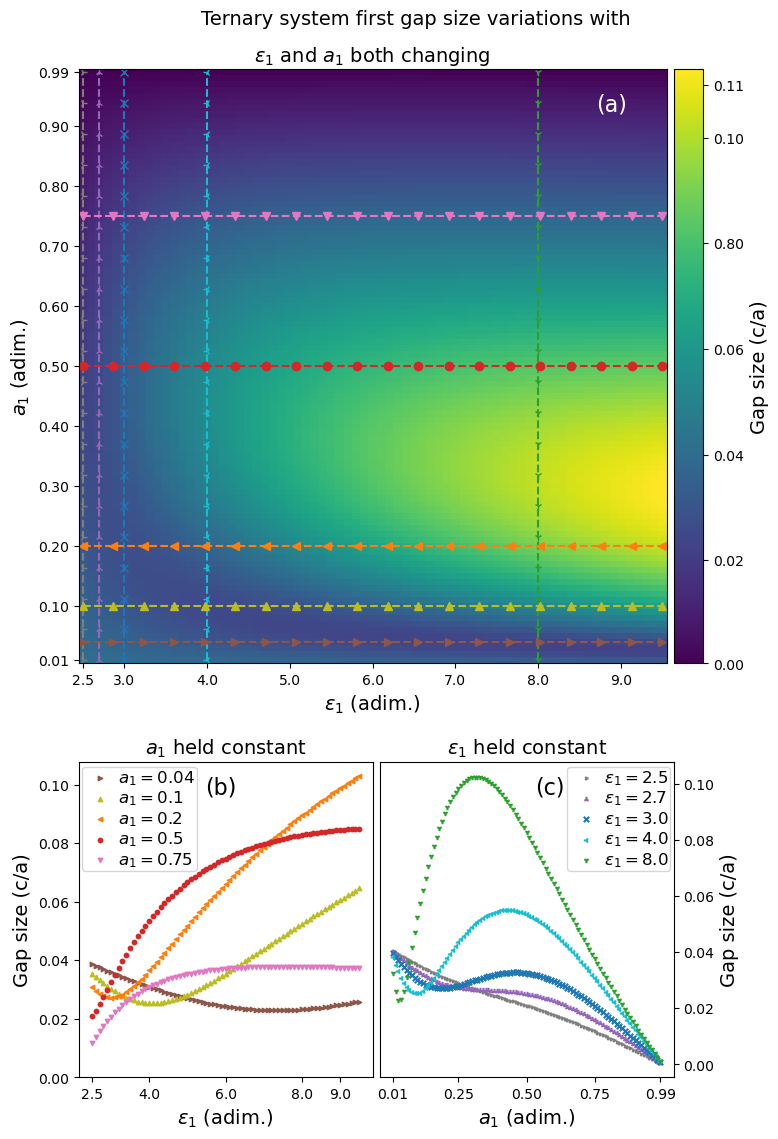}
    \caption{ Variations on a ternary system of the size of the first gap (a) including both $\epsilon_1$ and $a_1$ variations (b) varying $\epsilon_1$ but holding $a_1$ at a fixed value (c) varying $a_1$ but holding $\epsilon_1$ at a fixed value.}
    \label{fig:3mat-ea-gap1}
\end{figure}

\begin{figure}[ht!]
    \centering
    \includegraphics[scale=0.43]{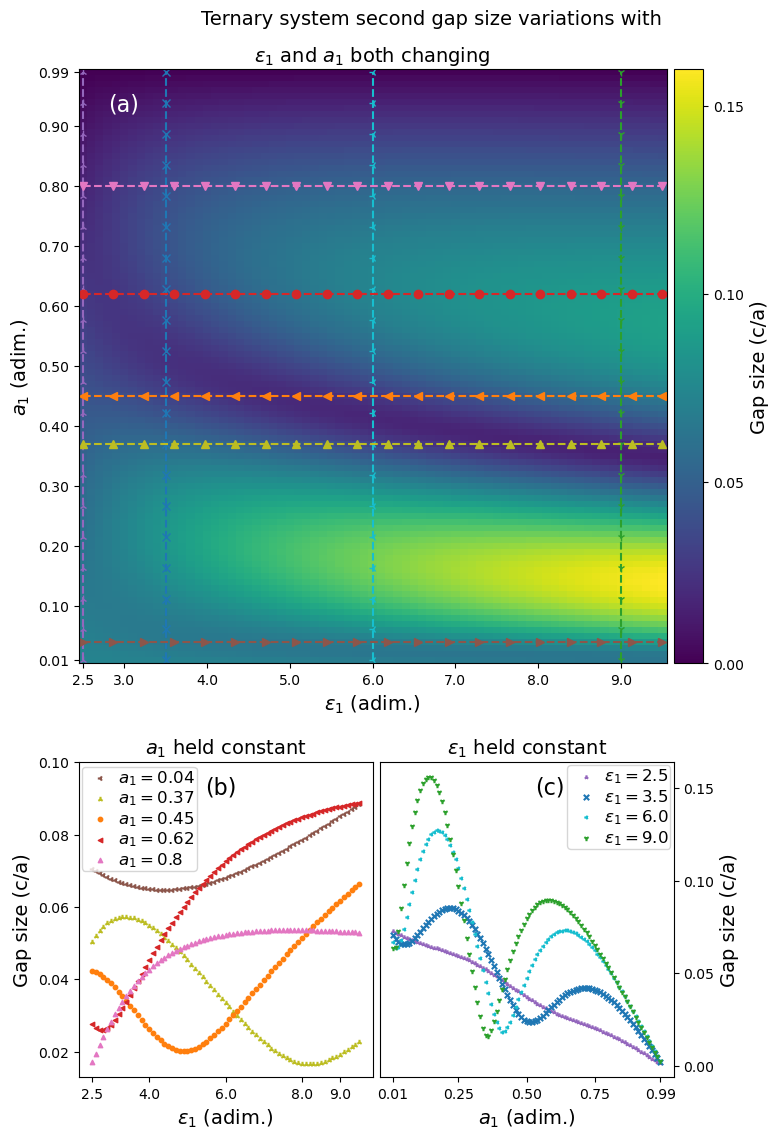}
    \caption{ Variations on a ternary system of the size of the second gap (a) including both $\epsilon_1$ and $a_1$ variations (b) varying $\epsilon_1$ but holding $a_1$ at a fixed value (c) varying $a_1$ but holding $\epsilon_1$ at a fixed value.}
    \label{fig:3mat-ea-gap2}
\end{figure}

The first group of calculations, varying the parameters $\epsilon_1$, $a_1$, is the most similar one to the one done for the binary system (shown in figures \ref{fig:3mat-ea-gap1}, \ref{fig:3mat-ea-gap2}). It refers to a situation where the size ($a_1$) and dielectric constant ($\epsilon_1$) values of the material with the highest dielectric constant are changed at will. The values of the other parameters where fixed to $\epsilon_2 = 2.4$, $a_2 = 4a_3$ (or $a_2 = 4(1-a_1)/5$). 

For the first gap (figure \ref{fig:3mat-ea-gap1}(a)), the behavior is very similar to the binary case, but an added region in the low $\epsilon_1$ and $a_1$ region with a significant gap size. The specific effects on the variations are more easily seen in figures \ref{fig:3mat-ea-gap1}(b) and \ref{fig:3mat-ea-gap1}(c). When varying $\epsilon_1$ while keeping $a_1$ constant (figure \ref{fig:3mat-ea-gap1}(b)), a ``new" moving local minima appears for low values of $a_1$ (figure \ref{fig:3mat-ea-gap1}(b) (yellow, orange, red)). The minima also moves towards lower values of $\epsilon_1$ as $a_1$ is increased. Afterwards, the behavior of the gap recovers the behavior of the same projection for a binary material (see \ref{fig:3mat-ea-gap1}(b) (pink, brown) and \ref{fig:2mat-gap1}(b)). On the other hand, when varying $a_1$ while keeping $\epsilon_1$ fixed, an even bigger difference in behavior emerges. First, for very small values of $\epsilon_1$ (figure \ref{fig:3mat-ea-gap1}(c) (purple)), the behavior is almost perfectly linear. Then, as $\epsilon_1$ is increased, the variations develop a saddle point (figure \ref{fig:3mat-ea-gap1}(c) (blue)), and then a local minima and maxima (figure \ref{fig:3mat-ea-gap1}(c) (cyan, green)). For higher values of $\epsilon_0$ the variations resemble more like those of a binary system (see figure \ref{fig:2mat-gap1}(c)). However, the local minima do not appear in the binary situation, which appear for all fixed values of $\epsilon_1$ above the saddle point.

To some extent, this situation can be interpreted as the effect of adding a third material on a binary system. Both cases, where $\epsilon_1 = 2.5$ and $a_1 = 0.01$ resemble binary configurations where, adding the third material would initially result in a smaller gap size (figures \ref{fig:3mat-ea-gap1}(b) (yellow, orange, red), (c)), but for large enough values of $a_1$ the gap size can increase even for slightly larger values of $\epsilon_1$ (figures \ref{fig:3mat-ea-gap1}(b) (pink, brown)). 

The green region in the lower left corner of the color map (figure \ref{fig:3mat-ea-gap1}(a) lower left) that qualitatively differentiates the binary (figure \ref{fig:2mat-gap1}(a)) and ternary cases (from what can be seen only from the color maps) can thus be identified with the fact that the same region corresponds to two different situations. In the binary case, the lower left region in figure \ref{fig:2mat-gap1}(a) corresponds to a material with very small differences from a regular, non-PC material with a single dielectric constant; or in this case, a material which is mostly composed by air (since $\epsilon_3=1$), for which there is no photonic band gap, or a very small one. However, that same region in a ternary system corresponds to introducing a third material into an already existing binary PC, with an already existing photonic band gap. That is why, the same region appears as having very small values (deep blue in color map) for the binary case, but sizeable ones (green in color map) for the ternary system. 

The overall effect of introducing a third material in the first gap seems to be making the local maxima smaller in scale (see color bars in figures \ref{fig:2mat-gap1}(a), \ref{fig:3mat-ea-gap1}(a)), but also introducing new ``moving" local minima in both projections. It is important to note that, whereas the local minima for the binary material reach a gap size of zero (up to the limit allowed by the resolution of the numerical calculations)[\cite{mpb}], the new minima do not destroy the gap.

For the second gap (figure \ref{fig:3mat-ea-gap2}), all the same visible effects for the first gap can be seen. The general behavior  seems very similar to the binary case (figure \ref{fig:3mat-ea-gap2}(a)), except for the existence of sizable gaps in the low $\epsilon_1$ and low $a_1$ region, just as in the case of the first gap (which also may be attributed to the fact that the binary $\epsilon_1=2.4$,  or $a_1=0$ already possesses a band gap). Specifically for the projections; varying $\epsilon_1$ while $a_1$ is held constant shows a similar new minima (figure \ref{fig:3mat-ea-gap2}(b) (yellow)), as in the case for the first gap, which moves as $a_1$ increases and then the behavior resembles that of figure \ref{fig:2mat-gap2}. However, when the projections cross the ``valley" in figure \ref{fig:3mat-ea-gap2}(a), the ``moving" minima obtained (figures \ref{fig:3mat-ea-gap2}(b) (orange, red, pink)) is ``lifted" (in the sense that it does not reach zero) compared to the one obtained for the binary case. Finally, the behavior again resembles that of the binary case for high enough values of $a_1$.

A similar effect is seen for the gap size variations for changes in $a_1$ keeping $\epsilon_1$ fixed. The behavior starts out close to linear, as in the case for the first gap, and resembles the shapes seen for the binary material as $\epsilon_1$ is increased, but the local minima seen in figure \ref{fig:2mat-gap2}(c) is now ``lifted".

In general, for all other gap numbers investigated, a similar effect can be seen: a number of ``hills" equal to the gap number and one less number of ``valleys", with the added effect that local minima associated with these ``valleys" are ``lifted", in the sense that the local minima for the ternary material do not reach values as low as the local minima for the binary material (figures \ref{fig:2mat-gap1}(c),\ref{fig:2mat-gap2}(b),\ref{fig:2mat-gap2}(c), \ref{fig:3mat-ea-gap1}(c), \ref{fig:3mat-ea-gap2}(b), \ref{fig:3mat-ea-gap2}(c)).

Now, for the second group of calculations, varying $\epsilon_1$ and $\epsilon_2$, the purpose was to understand the effect of varying $\epsilon2$ while still having a connection to the already seen variations of $\epsilon_1$. However, since $\epsilon_2$ is constrained by $\epsilon_1>\epsilon_2$ by convention, it is more convenient to vary $\epsilon_2$ indirectly through a parameter $0 \leq \eta_{\epsilon} \leq 1$ that ensures that the entire range of possible values of $\epsilon_2$ is covered for every value of $\epsilon_1$. The formula relating $\epsilon_2$ and $\eta_{\epsilon}$ is
\begin{equation}
    \epsilon_2 = \epsilon_1 \eta_{\epsilon} +\epsilon_3 (1-\eta_{\epsilon}) \rightarrow \epsilon_2 = \epsilon_1 \eta_{\epsilon} + (1-\eta_{\epsilon}) \quad ,
\end{equation}
the last expression is for $\epsilon_3=1$, and from this it is possible to obtain
\begin{equation}
    \eta_{\epsilon} = \frac{\epsilon_2-\epsilon_3}{\epsilon_1-\epsilon_3} = \frac{\epsilon_2-1}{\epsilon_1-1} \quad .
\end{equation}

Thus, $\eta_{\epsilon}(\epsilon_2)$ is a linear function (in $\epsilon_2$) that goes from 0 to 1 as $\epsilon_2$ goes from $\epsilon_3$ to $\epsilon_1$. Therefore, the cases $\eta_{\epsilon}=0$ and $\eta_{\epsilon}= 1$ are actually binary materials, a fact that helps in the interpretation of the results. The relative sizes of the materials were kept fixed. $a_2$ and $a_3$ retain the relation $a_2=4a_3$, and $a_1$ was fixed at $a_1=0.4$; then, $a_2=0.48$ and $a_3=0.12$.

\begin{figure}[ht!]
    \centering
    \includegraphics[scale=0.43]{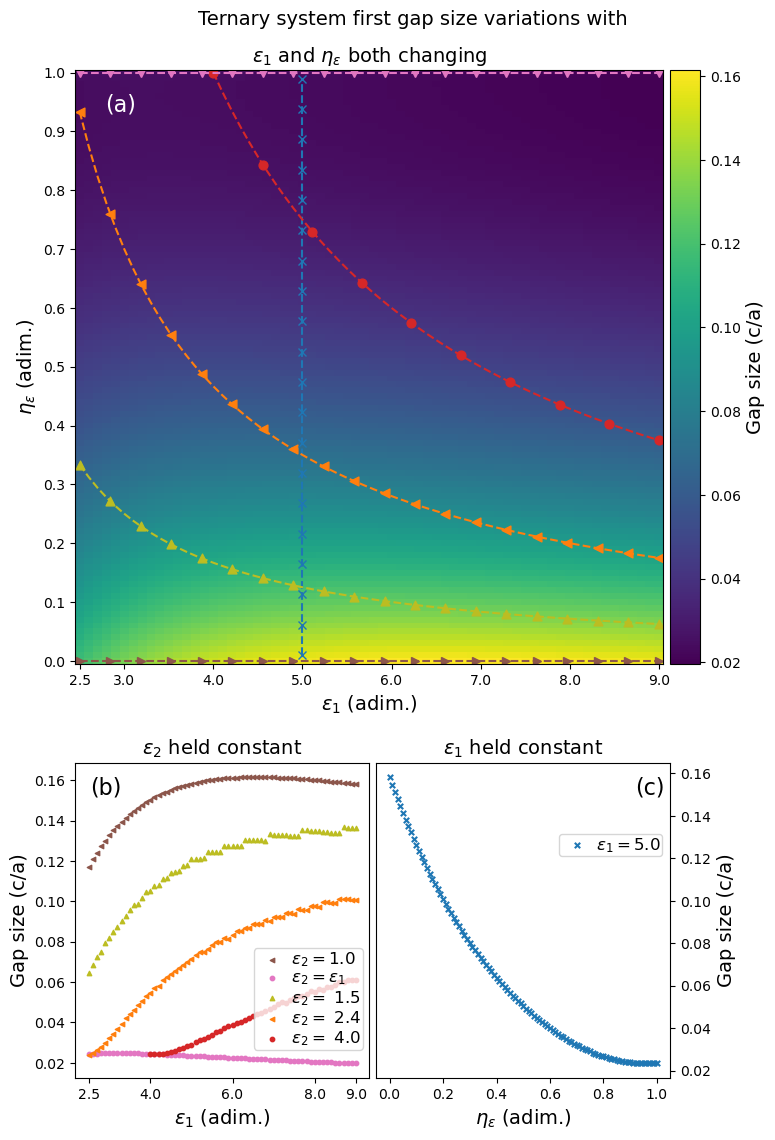}
    \caption{Variations on a ternary system of the size of the first gap (a) including both $\epsilon_1$ and $\eta_{\epsilon}$ variations, (b) varying $\epsilon_1$ but holding $\eta_{\epsilon}$ in a fixed value, (c) varying $\eta_{\epsilon}$ but holding $\epsilon_1$ in a fixed value. }
    \label{fig:3mat-ee-gap1}
\end{figure}

For the first gap, there is a clear decreasing behavior of the gap size for increasing $\eta_\epsilon$, as seen in figures \ref{fig:3mat-ee-gap1}(a),(c). This implies that adding a third material may have the effect of increasing or decreasing the gap size depending on whether the dielectric constant of the third material is closer to the the lower dielectric constant material or to the higher one. The two limiting binary systems, $\eta_\epsilon=0$ and $\eta_\epsilon=1$, are shown in brown and pink respectively in figure \ref{fig:3mat-ee-gap1}(b). Additionally, the variations for $\epsilon_2 = 2.4$ are shown in figure \ref{fig:3mat-ee-gap1}(b)(orange), which corresponds to a variation calculated in the first group of calculations. Both cases allow us to see the effect of varying the $\epsilon_2$ parameter: the variations retain the behavior seen for the binary case and the first group of ternary calculations as in figures \ref{fig:2mat-gap1}(b),\ref{fig:3mat-ee-gap1} for all values of $\epsilon_2$; but as $\epsilon_2$ changes, the scale of the variations changes too.

\begin{figure}[ht!]
    \centering
    \includegraphics[scale=0.43]{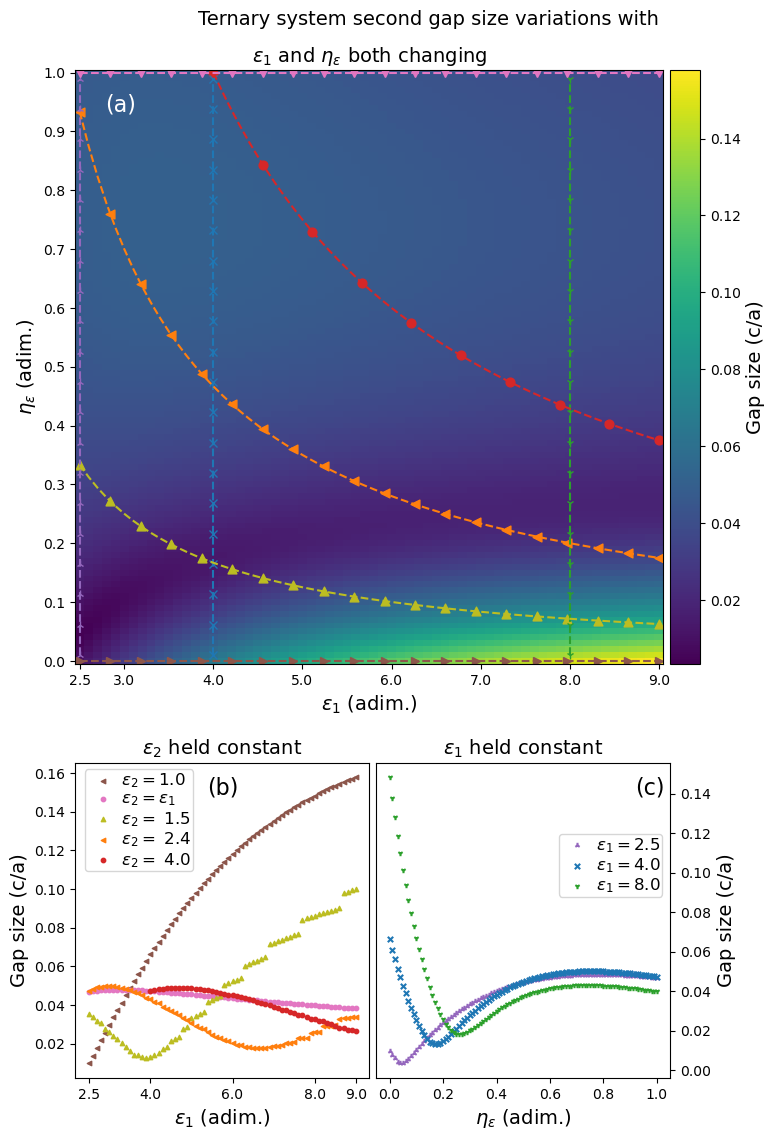}
    \caption{Variations on a ternary system of the size of the second gap (a) including both $\epsilon_0$ and $\eta_{\epsilon}$ variations, (b) varying $\epsilon_0$ but holding $\eta_{\epsilon}$ in a fixed value, (c) varying $\eta_{\epsilon}$ but holding $\epsilon_0$ in a fixed value. }
    \label{fig:3mat-ee-gap2}
\end{figure}

The effects for the second gap can be seen in figure \ref{fig:3mat-ee-gap2}. The main additional feature for the second gap is the appearance of a ``valley" (see figure \ref{fig:3mat-ee-gap2}(a)) in the gap size variations. This is reflected in the variations, keeping $\epsilon_1$ fixed. When the value of $\epsilon_2$ is increased, the gap size reaches a local minima very fast (see figure \ref{fig:3mat-ee-gap2}(c)), but then reaches a local maxima with a more ``plateau" like behavior as it approaches $\epsilon_1$. On the other hand, when $\epsilon_2$ is held constant (figure \ref{fig:3mat-ee-gap2}(b)), in addition to the scaling effect that is seen for the first gap, varying $\epsilon_2$ also has an effect in the position of the local minima characteristic of the $\epsilon_1$ variations. Increasing $\epsilon_2$ also increases the location of the local minima. For the first group of calculations this local minima correspond to intersections between $a_1=cte$ curves and ``valleys"; so increasing $\epsilon_2$ also raises the ``valleys" for variations like the first group of variations.

\begin{figure}[ht!]
    \centering
    \includegraphics[scale=0.48]{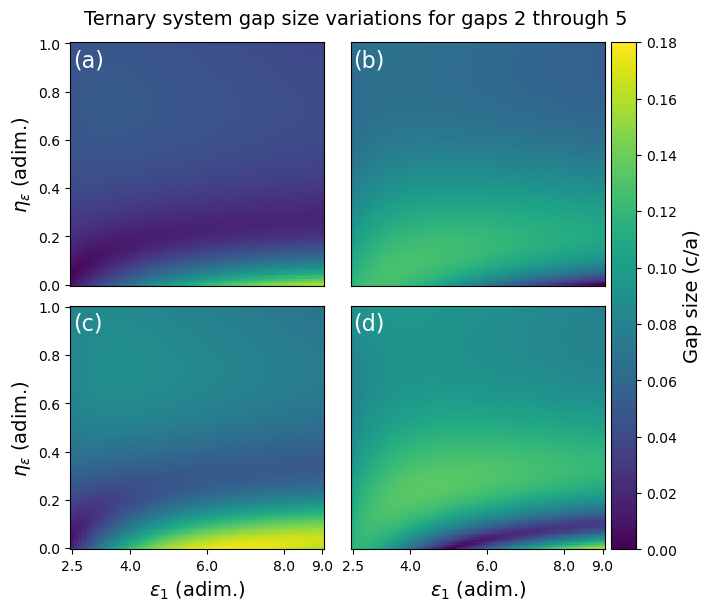}
    \caption{  Variations on a ternary system of the size of the (a) second, (b) third, (c) fourth, (d) fifth gaps when both $\epsilon_1$ and $\eta_{\epsilon}$ are varied.}
    \label{fig:3mat-ee-4gaps}
\end{figure}

Furthermore, the effect on the 3rd gap seems to be the opposite than for the second gap, having a sharp local maxima for low $\eta_{\epsilon}$ and settling into a plateau as $\eta_{\epsilon}$ increases figure (\ref{fig:3mat-ee-4gaps}(b)).

In general, the gaps studied seem to be able to be divided into three groups (figure \ref{fig:3mat-ee-4gaps}): the first gap (or type 0), which is the only one with a strictly monotonically decreasing behavior for increasing $\eta_{\epsilon}$, the even gaps (or type -), whose main feature is a rapidly appearing local minima that then grows to a local maxima as $\eta_{\epsilon}$ is increased (figures \ref{fig:3mat-ee-gap2}(c), \ref{fig:3mat-ee-4gaps}(a),(c)), and the odd gaps different from the first gap (or just odd gaps or type -), whose behavior seems to be opposite in many ways to the even gaps, first having a rapid increase to a sharp local maxima and then settling into a plateau (figure \ref{fig:3mat-ee-4gaps}(b),(d)). However, in gap 5 (figure \ref{fig:3mat-ee-4gaps}(d), an odd gap) a small local maxima appears for small values of $\eta_{\epsilon}$, something that defies the behavior expected from the other gaps (2-4). In fact, looking at even higher gap numbers, the behavior of even gaps resembles that of odd gaps for high values of $\epsilon_1$, and the behavior of odd gaps resembles that of even gaps for high values of $\epsilon_1$. For high enough gap number, the behavior can switch back again to the original one for high values of $\epsilon_1$. Thus, the general behavior of the even (odd) gaps is to flip between type - and type + behavior starting with type - (+) behavior. Determining the number of flips that a gap can have in a given range for $\epsilon_1$ requires further investigation to fully understand.

The final group of calculations explores changes in $a_2$ while keeping a connection with the results in the first group of calculations ($\epsilon_1$, $a_1$) through the variations in $a_1$. Similarly to the second group of calculations, an auxiliary parameter $\eta_a$ was used to cover the possible range of values for $a_2$ for every value of $a_1$ chosen. In particular, the equation defining $\eta_a$ is 
\begin{equation}
    \eta_a = \frac{a_2}{1-a_1} \quad .
\end{equation}

It comes from the normalization condition for the sizes of the materials in the ternary system
\begin{equation}
    a_1 + a_2 + a_3 = 1 \rightarrow a_2 + a_3 = 1 - a_1 \quad .
\end{equation}
Thus $\eta_a$ can be interpreted as the filling fraction of the material with dielectric constant $\epsilon_2$ of the space left after fixing $a_1$.
\begin{equation}
    \frac{a_2}{1-a_1} + \frac{a_3}{1-a_1} = 1 \quad .
    \label{ec:renormalizado}
\end{equation}

For this group of calculations, it is also true that changes in $\eta_a$ involve changes in both $a_2$ and $a_1$, but that is already the case for all of the calculations previously considered; for the binary system, changing $a_1$ also changes $a_2$, and for the ternary system (the first group of calculations), we chose to keep the ratio between $a_2$ and $a_3$ constant as $a_1$ varied. This sort of variation is the one that a fixed value of $\eta_a$ considers, we thus focus again in fixed values of both parameters $(a_1,\eta_a)$. 

\begin{figure}[ht!]
    \centering
    \includegraphics[scale=0.43]{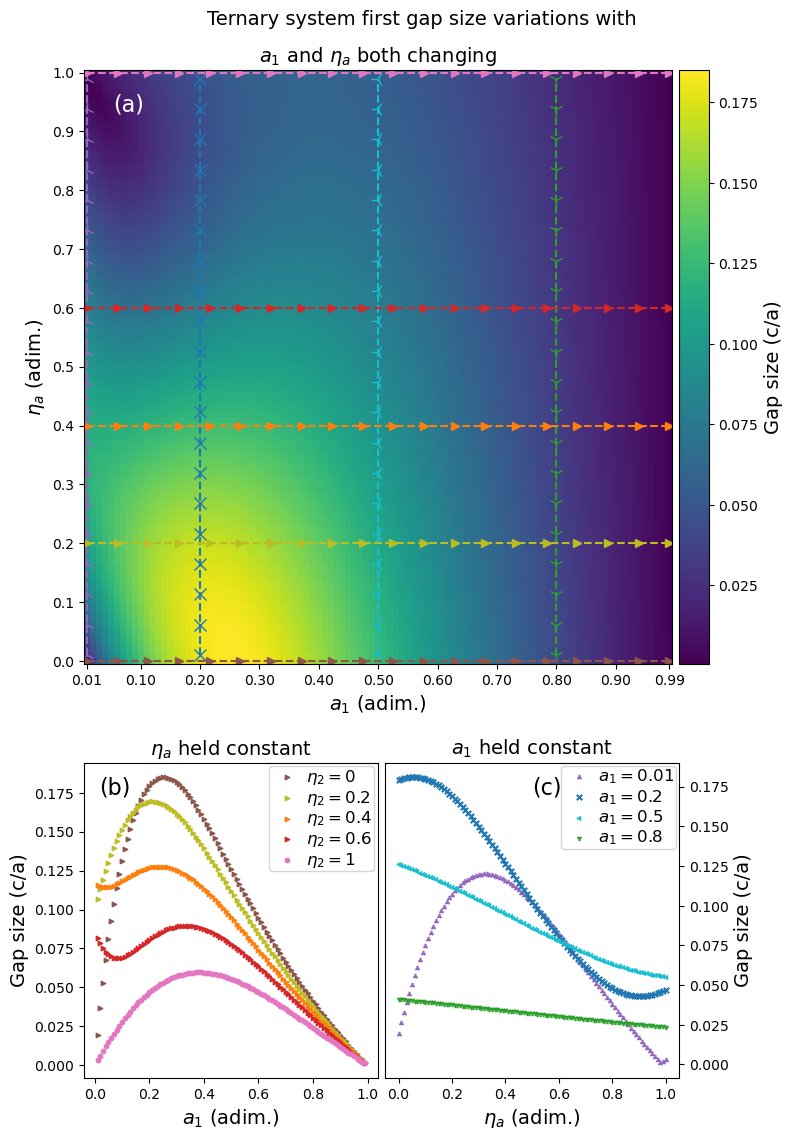}
    \caption{Variations on a ternary system of the size of the first gap (a) including both $\eta_a$ and $a_1$ variations, (b) varying $a_1$ but holding $\eta_a$ in a fixed value, (c) varying $\eta_a$ but holding $a_1$ in a fixed value. }
    \label{fig:3mat-aa-gap1}
\end{figure}

For the first gap, the variations have a single ``hill" (figure \ref{fig:3mat-aa-gap1}(a)). In the binary cases ($\eta_a=0$, $\eta_a=1$), shown in figure \ref{fig:3mat-ea-gap1}(b)(brown, pink), we can see that the rest of $\eta_a$ values recover the same qualitative behavior that is shown in the first group of calculations for the ternary system (figure \ref{fig:3mat-ea-gap1}(c)). Thus the appearance of the local minima is thus expected for a wide variety of ternary systems, yet some values of $\eta_a$ do not have a local minima (e.g $\eta_a=0.2$ figure \ref{fig:3mat-aa-gap1}(b)(yellow)). Varying $\eta_a$ would also have an effect on the overall scale of the variations, just as varying $\epsilon_2$ did in the second group of calculations.

On the other hand, when $a_1$ is fixed we see that for low values of $a_1$ fixed, the behavior is very similar to a binary one (\ref{fig:3mat-aa-gap1}(c) (purple)), which makes sense since one of the materials ($a_1$ in this case) does not represent a significant portion of the ternary system. As $a_1$ is increased, the behavior tends towards a linear decreasing one (\ref{fig:3mat-aa-gap1}(c) (blue, cyan, green)), it again makes sense that at high $a_1$ the behavior is more monotonous since is now $a_2$ the one that has a small portion in the ternary (\ref{fig:3mat-aa-gap1}(c) (green)). The intermediate values of $a_1$ (\ref{fig:3mat-aa-gap1}(c) (blue, cyan)) seem to take the behavior from the binary one to the linear one by ``peeling" the low $\eta_a$ behavior of the smaller $a_1$ fixed variations. Notice that for $a_2=0.2$ (\ref{fig:3mat-aa-gap1}(c) (blue)), there is no longer a rise to the local maxima seen in (\ref{fig:3mat-aa-gap1}(c) (purple)) and in (\ref{fig:3mat-aa-gap1}(c) (cyan)); the inflexion point at the local maxima is gone, and the behavior is already more similar to the linear one than to the binary one. This behavior is better exemplified in the second gap.

\begin{figure}[ht!]
    \centering
    \includegraphics[scale=0.43]{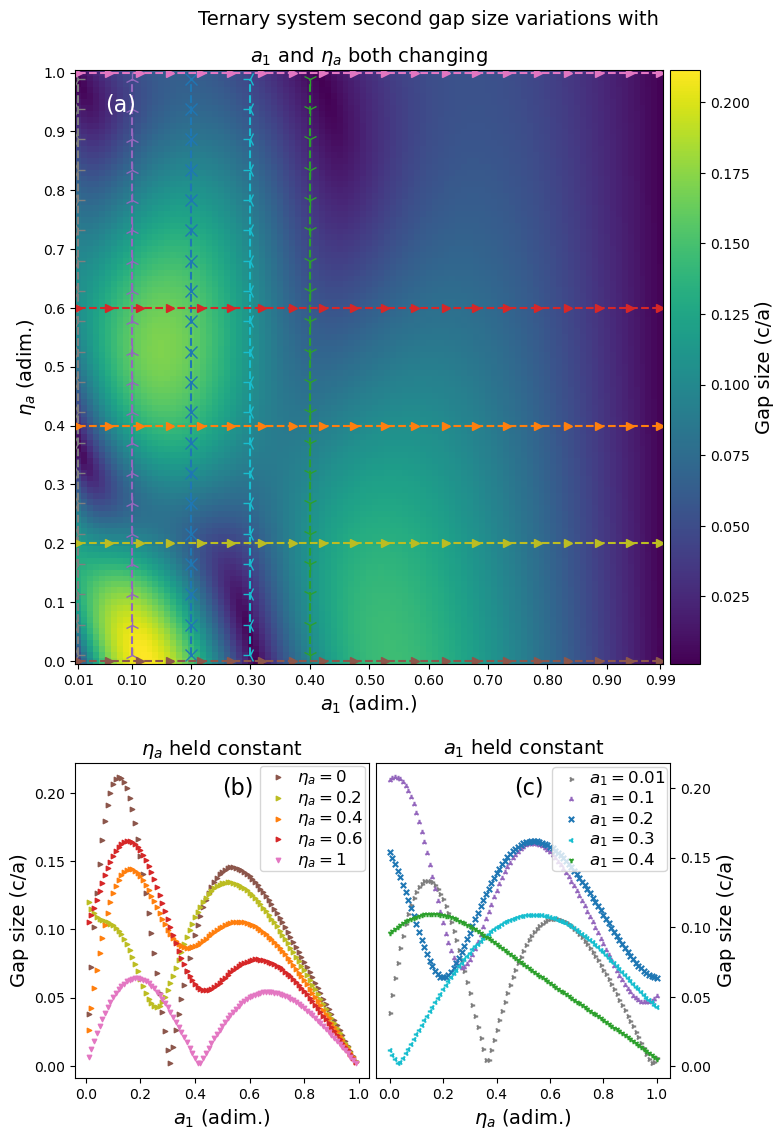}
    \caption{Variations on a ternary system of the size of the second gap (a) including both $\eta_a$ and $a_1$ variations, (b) varying $a_1$ but holding $\eta_a$ in a fixed value, (c) varying $\eta_a$ but holding $a_1$ in a fixed value. }
    \label{fig:3mat-aa-gap2}
\end{figure}

For the second gap, the variations are shown in figure \ref{fig:3mat-aa-gap2}. The general features show three ``hills", two of them are connected to the one in the upper left but not among each other. Additionally, when $\eta_a$ is fixed (figure \ref{fig:3mat-aa-gap1}(b)), the resulting variations show a more complex behavior as $\eta_a$ is fixed at different values. The scaling effect seen for the first gap only applies in a simple fashion to the ``hill" corresponding to higher $a_1$ values, the other ``hill" may even disappear for some $\eta_a$ values (e.g $\eta_a=0.2$). The effect that changing $\eta_a$ has on the position of the ``valleys" (in variations of the type ($\epsilon_1$, $a_1$)) is also more complex than the one that $\epsilon_2$ has. Whereas $\epsilon_2$ moves the ``valleys" upward, $\eta_a$ moves the ``valley" in a ``zig zag" fashion.

When $a_1$ is held constant the effect is as described for the first gap. For low values (\ref{fig:3mat-aa-gap2}(c) (purple)), the behavior is very similar to the binary one, but as $a_1$ is increased the same "peeling" effect takes place. First, the rise to the local maxima is no longer visible (\ref{fig:3mat-aa-gap2}(c) (blue)), then the inflexion point vanishes (\ref{fig:3mat-aa-gap2}(c) (cyan)); for larger values of $a_1$ the drop to the local minima also disappears (\ref{fig:3mat-aa-gap2}(c) (green)); and for larger values of $a_1$ the rise towards the second maxima is also peeled from the variations (\ref{fig:3mat-aa-gap2}(c) (gray)). 

\begin{figure}[ht!]
    \centering
    \includegraphics[scale=0.47]{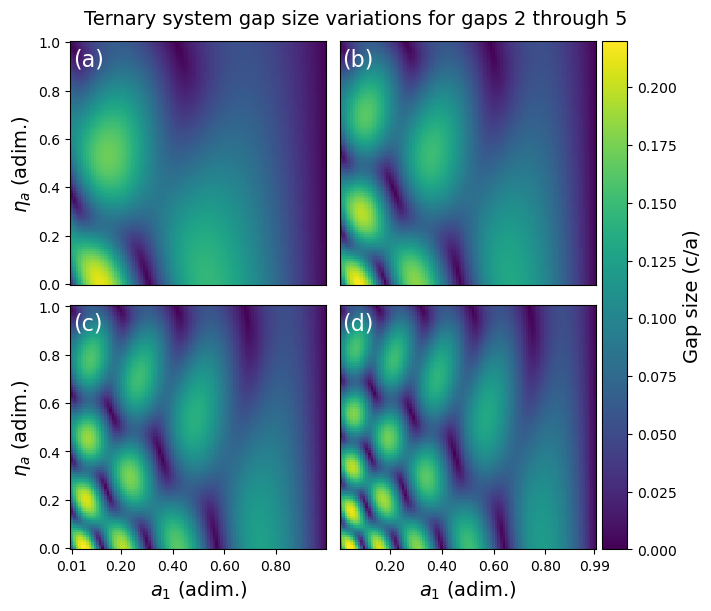}
    \caption{ Variations on a ternary system of the size of the (a) second, (b) third, (c) fourth, (d) fifth gaps when both $a_1$ and $\eta_{a}$ are varied.}
    \label{fig:3mat-aa-4gaps}
\end{figure}

Thus, the general behavior for any arbitrary gap (figure \ref{fig:3mat-aa-4gaps}) seems to be a number of ``hills" equal to the gap number for low $a_1$, but as $a_1$ is increased, the ``hills" are ``peeled" away and leaving only a linearly decreasing behavior (in $\eta_a$) for large $a_1$.

For the ternary system, the gap position has a similar regular behavior than the binary one. We again compared the Drude and OPL EMA flavours with the numerical data by adding a third term in (\ref{ec:drude}),(\ref{ec:paralelo}). We found very similar results to those for the binary case. The OPL flavour gives a better estimate overall but Drude does for systems that are similar to homogeneous systems. The OPL flavour actually gave a slightly lower maximum relative error than the binary case for the configurations studied here. 

On the other hand, no generalization of the formulas for ``valleys" (\ref{ec:mu-valles}), ``ridges" (\ref{ec:mu-crestas}) or gap percentage estimation (\ref{ec:ext-formula}) were found. In any case, the qualitative similarities between the binary and ternary cases do imply that a formula could be given to estimate the gap percentage of ternary systems.

Overall, ternary systems show a similar qualitative behavior than binary ones. However, our results suggest that the binary system is capable of slightly larger gap sizes. Thus, a binary system may suffice for most applications but, since ternary systems allow to control the positions of local minima and local maxima, ternary systems may allow for additional flexibility if needed.

\subsection{Linear dielectric function and limiting behavior of n-layer material}

\begin{figure}[ht!]
    \centering
    \includegraphics[scale=0.45]{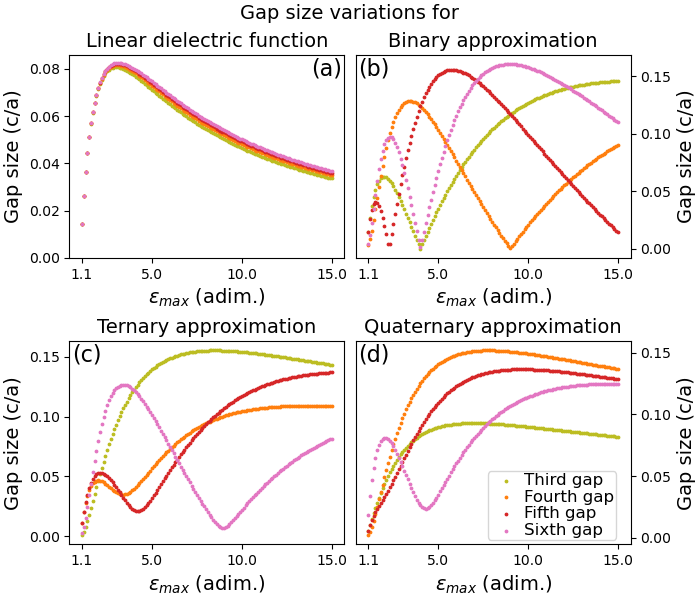}
    \caption{ Gap size variations for the third through sixth gaps for (a) the linear dielectric function material, (b) binary fixed ends approximation, (c) ternary fixed ends approximation, (d) quaternary fixed ends approximation.}
    \label{fig:lineal}
\end{figure}

For materials with a linear dielectric function, there is only one parameter we are interested in varying, $\epsilon_{max}$. The behavior of the gap size for different $\epsilon_{max}$ can be seen in figure \ref{fig:lineal}(a). All gaps show a similar behavior, which resembles the behavior of the gap size while varying $\epsilon_1$ at fixed high $a_1$ (figure \ref{fig:2mat-gap1}(b) (red)), but with a steeper drop after the local maxima. Nevertheless, it is intriguing that all the increasingly  complex gap size behavior observed for the binary and ternary systems, with multiple appearing local minima and thus also additional local maxima, is not reflected in the system with linear dielectric function. Indeed, even the unique behaviors of each individual gap are erased in the linear case. 

\begin{figure}[ht!]
    \centering
    \includegraphics[scale=0.4]{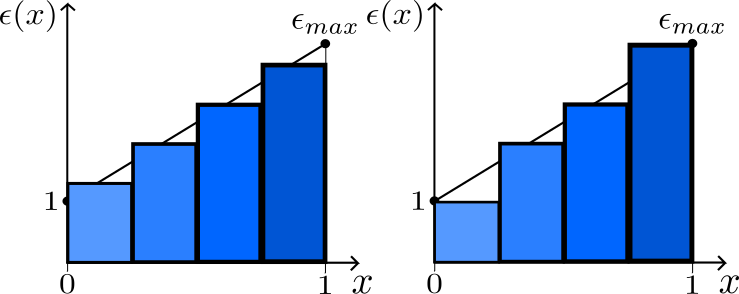}
    \caption{ Comparison of the two approximation methods considered: straightforward approximation (left), fixed ends approximation (right). }
    \label{fig:aprox-lineal}
\end{figure}

To give a better understanding of the underlying phenomena, we devised additional calculations which aimed to simulate the process of a multi-layer material approaching a linear one. Initially, the simplest approximation consisted in approximating the linear material with a n-layer material in which all of the layers shared the same size, and their respective $\epsilon_i$ were determined by the value of the linear dielectric function at the centers of each layer (figure \ref{fig:aprox-lineal}). Although this does tend towards the linear case, a second approximation with more interesting features emerged by enforcing the values $\epsilon_n=1$ and $\epsilon_1=\epsilon_{max}$. Both approximation processes have the same limiting behavior, and the mechanisms which drive the multi-layer towards the linear  dielectric function behavior are the same in both cases. The only reason we chose to showcase the approximation process with fixed ends was because of the greater number of local minima in the variations, which provided better opportunities to see the limiting behavior in action. We note that the first approximation method approaches the regular behavior faster.

Figures \ref{fig:lineal}(b,c,d) show the gap size variations for the binary, ternary and quaternary approximations of the linear dielectric function described above. The gaps included were the third through sixth, since the first gap has variations similar to figure \ref{fig:2mat-gap1}(c) (red) all of the n-layer approximations, and thus, it does not add valuable information since it already relates to the linear case. On the other hand, the binary approximation \ref{fig:lineal}(b) has local minima for all gaps shown, in a stark contrast to the linear case. One of the mechanisms that turns the binary behavior into the linear one is the already mentioned local minima ``lifting" seen in the previous section. Indeed, figure \ref{fig:lineal}(c) shows the local minima ``lifting" of three of the local minima in the binary case. Furthermore, the variations for the second gap, already show a single local maxima. Already in the four-layer case, only one of the gaps still shows a ``lifted" local minima, all other gaps behave similar to the limiting behavior.

Therefore, the very stable properties of the gap size for linear function variations can be approximated with multi-layer materials, but the number of materials required depends on the gap number in question. The higher the gap number, the more materials will be needed to smooth out the behavior.

We may intuitively generalize the EMA approach to linear dielectric function systems, by substituting the sum of a finite number of terms by an integral over the position in the primitive cell of the PC. The resulting estimate is similar for both the Drude and OPL flavours but differ drastically from the numerical data. One way to fix the problem is to assume that the gap position is given by the formula
\begin{equation}
    \label{ec:efectivo-continuo}
    \overline{\omega} = \frac{m}{2 \epsilon_{eff}} = \frac{m}{ \epsilon_{max}+1}  \quad ,
\end{equation}
rather than (\ref{ec:efectivo}) (the second equality corresponds to the continuum Drude flavour estimation).

Furthermore, the Drude flavour shows a better fit to the data, contrary to the results in the rest of the calculations, where the OPL flavour yielded better results. This could suggest that the Drude EMA flavour could be favored when estimating gap positions of dielectric functions are defined by continuous gratings.

\section{\label{sec:conclus}Conclusions}

Binary systems ``band gap atlas" show a regular pattern that allowed a clear qualitative and quantitative description. The utility of these results lies not only in the qualitative description of the photonic band gap dynamics and their structure but also on the possibility to estimate both gap size and position using simple formulas, which allows to explore a wide range of possible designs with minimal computational cost. This advantage is greatest for cases where dynamic control is wanted since the behavior for all the configurations of the system that would be used need to be accounted for. 

Also, ternary systems show a behavior similar than the binary case. The emerging phenomena and role of each of the parameters has on the gap size is qualitatively clear but quantitatively complex. The qualitative results do nevertheless give crucial information on the effect that adding a third material material may have on the gap size of a dielectric structure. Overall, this third material does not seem to give any drastically different behavior to the binary option, but does offer the possibility of added flexibility if needed.

Finally, the linear grating system shows a regularity behavior that was able to be reproduced using only a limited number of materials, something that at least partially explains the way that his regularity emerges from the subsequent addition of extra materials but also may put in reach the use of this kind of regularity in applications where the use of a dielectric grating may be difficult. The behavior of the gap position showed an anomalous behavior that defied estimation with the use of the EMA models that where successful in the binary and ternary cases. The gap position was described using an alternative expression for the general effective medium gap estimation formula. This alternative formula may also prove better at estimating the gap position of other dielectric grating systems.

\begin{acknowledgments}
We are thankful for the financial support from Universidad Nacional de Colombia through the 57522 research project.
Figures were generated using the Inkscape\cite{Inkscape} software and the Matplotlib\cite{matplotlib} Python package. Analysis made strong use of NumPy\cite{numpy} and SciPy\cite{SciPy} packages.
\end{acknowledgments}

\section*{Conflict of Interest}
The authors declare  that there is no conflict of interest regarding the publication of this paper.

\section*{Authors Contributions}
O.D.H.P. developed the simulations, and wrote the manuscript. R.R.R.G. conceived the idea, revised the manuscript and supervised the project.

%\bibliography{export}
%
\end{document}